\documentclass[%
 aip,
 amsmath,amssymb,
 reprint,%
]{revtex4-1}
\usepackage{graphicx}% Include figure files
\usepackage{dcolumn}% Align table columns on decimal point
\usepackage{bm}% bold math
\usepackage[utf8]{inputenc}
\usepackage[T1]{fontenc}
\usepackage{mathptmx}
\usepackage{etoolbox}

\makeatletter
\def\@email#1#2{%
 \endgroup
 \patchcmd{\titleblock@produce}
  {\frontmatter@RRAPformat}
  {\frontmatter@RRAPformat{\produce@RRAP{*#1\href{mailto:#2}{#2}}}\frontmatter@RRAPformat}
  {}{}
}%
\makeatother

\begin{document}

\preprint{AIP/123-QED}

\title{Facilitating {\it ab initio} configurational sampling of multicomponent
solids using an on-lattice neural network model and active learning}
\author{Shusuke Kasamatsu}
\email{kasamatsu@sci.kj.yamagata-u.ac.jp}
\affiliation{Academic Assembly (Faculty of Science), Yamagata University, 1-4-12 Kojirakawa, Yamagata-shi,
Yamagata 990-8560 JAPAN}

\author{Yuichi Motoyama}
\affiliation{The Institute for Solid State Physics, the University  of Tokyo, 5-1-5 Kashiwanoha, Kashiwa-shi, Chiba, 277-8581, JAPAN.}

\author{Kazuyoshi Yoshimi}
\affiliation{The Institute for Solid State Physics, the University  of Tokyo, 5-1-5 Kashiwanoha, Kashiwa-shi, Chiba, 277-8581, JAPAN.}

\author{Ushio Matsumoto}
\affiliation{Nanostructures Research Laboratory, Japan Fine Ceramics Center, 2-4-1 Mutsuno, Atsuta-ku, Nagoya 456-8587 JAPAN.}
\affiliation{Department of Materials Science and Engineering, Kyoto University, Yoshida-Honmachi, Sakyo-ku, Kyoto 606-8501, JAPAN.
}
\author{Akihide Kuwabara}
\affiliation{Nanostructures Research Laboratory, Japan Fine Ceramics Center, 2-4-1 Mutsuno, Atsuta-ku, Nagoya 456-8587 JAPAN.}

\author{Takafumi Ogawa}
\email{t\_ogawa@jfcc.or.jp}
\affiliation{Nanostructures Research Laboratory, Japan Fine Ceramics Center, 2-4-1 Mutsuno, Atsuta-ku, Nagoya 456-8587 JAPAN.}

\date{\today}% It is always \today, today,

\begin{abstract}
We propose a scheme for {\it ab initio} configurational sampling in multicomponent
crystalline solids using Behler-Parinello type neural network potentials (NNPs)
in an unconventional way:  
the NNPs are trained to predict the energies of relaxed structures from the perfect
lattice with configurational disorder instead of the usual way of training to
predict energies as functions of continuous atom coordinates. An active learning scheme
is employed to obtain a training set containing configurations of thermodynamic relevance. 
This enables bypassing of the
structural relaxation procedure which is necessary when applying conventional NNP approaches
to the lattice configuration problem. The idea is demonstrated on the
calculation of the temperature dependence of the degree of A/B site inversion in 
three spinel oxides, MgAl$_2$O$_4$, ZnAl$_2$O$_4$, and MgGa$_2$O$_4$.
The present scheme may serve as an alternative to cluster expansion for
`difficult' systems, e.g., complex bulk or interface systems with many
components and sublattices that are relevant to many technological applications today.
\end{abstract}

\maketitle
\section{Introduction}
Configurational order/disorder determines many properties of functional
materials including mechanical strength, catalytic activity, ion/electron/phonon
conductivity, and so on. Configurational disorder can be predicted using Monte
Carlo (MC) methods based on statistical mechanics, but MC methods usually
require a huge number of energy evaluations. Because of this, most works in the
past have relied on effective low-cost models that are fitted to density
functional theory (DFT) calculations \cite{Hafner2006,Ceder2010,neugebauer2013}. 
For solid-state systems that can be mapped onto a lattice, the cluster expansion
method \cite{Sanchez1984,Ceder1993, Walle2002,Seko2009,Sanchez2010,Sanchez2017, Wu2016, Chang2019} is often considered the {\it de facto} standard for obtaining such
effective models. This method describes configuration energies using an
expansion based on effective cluster interactions (ECIs) :
\begin{equation}\label{eq:CE}
    E(\vec{\sigma})  = \sum_\alpha m_\alpha V_\alpha  \Phi_\alpha (\vec{\sigma}),
\end{equation}
where $\alpha$ denotes empty, single, and multi-component clusters, 
$m_\alpha$ is the cluster multiplicity, $V_\alpha$ is
the ECI, and $\vec{\sigma}$ is the occupation vector. $\Phi_\alpha$ is the
cluster basis function, which for a binary system takes the simple form of
\begin{equation}
    \Phi_\alpha (\vec{\sigma}) = \prod_{p \in \alpha} \sigma_p,
\end{equation}
where $p$ are site indices and $\sigma_p = \pm 1$ depending on which component
occupies site $p$. Appropriate basis functions for systems with more components can
also be generated according to, e.g., Ref.~\citenum{Sanchez2010}.
We note that the validity and formal approximations involved in mapping a
realistic alloy model with vibrations and intra-atomic excitations to an on-lattice
model has been explored rather thoroughly by Ceder \cite{Ceder1993}.

The ECIs are usually fitted to energies of relaxed structures from small-scale
DFT calculations, then used to calculate energies of a much larger supercell in
the course of MC sampling calculations. Although this cluster expansion approach
has seen much success, especially for metallic alloys, certain limitations and
difficulties have also been pointed out over the years. For example, it is known
that the
accuracy of the cluster expansion energy prediction degrades when
there is significant lattice relaxation \cite{Nguyen2017}. In fact, cluster expansion with
constant ECIs is, in a way, a zeroth order approximation and formally cannot
provide an exact fitting including relaxation effects \cite{Sanchez2017}. Another difficulty is in
choosing the finite number of clusters that will give the best predictions.
For example,
Seko {\it et al.}~reported a spurious prediction of a discontinous order-disorder
transition in an oxide
system with long-range interactions when choosing clusters that perform
best within limited supercell sizes \cite{Seko2014}.
Still another issue is in how to avoid undesirable training set bias. These issues are, in
fact, highly nontrivial, and developing robust methods for
cluster and training set selection is still an area of active research 
\cite{Seko2009,Seko2011,Seko2014,Chang2019,Leong2019}.  Also, cluster expansion
becomes computationally demanding in a combinatorial way as the
number of components and sublattices increases, and this limits the number of clusters
that can be included in the expansion \cite{Wu2016}. 

In view of these issues, we as well as some other workers have opted to bypass
fitted models and sample directly on DFT energies
\cite{Khan2016,Wexler2019,Kasamatsu2019,Kasamatsu2020}. Using more sophisticated
MC schemes that are suited for parallel computation such as
Wang-Landau \cite{Wang2001} or replica exchange Monte Carlo (RXMC)
\cite{Hukushima1996} sampling, sufficient statistical sampling has been achieved
on calculation models with up to a few hundred atoms. However, some of these
works required weeks of calculations on up to 100 supercomputer nodes, and much
acceleration is necessary if this type of approach is to be used in a more
widespread manner on a variety of materials systems. Towards this end, we
consider, in this work, the so-called ``active learning'' approach, which has
been developed to accelerate first-principles molecular dynamics simulations or
MC calculations
\cite{Podryabinkin2017,Podryabinkin2019,Gubaev2019,Smith2018,Zhang2019,Kostiuchenko2019,Loeffler2020,Vandermause2020,jinnouchi2020}.
The basic idea is to use machine-learning potentials (MLPs) that have been
fitted to DFT results to accelerate the calculations. Since MLPs are usually
good at interpolating but not at extrapolating, a relearning is performed when
the system wanders into a previously unlearned region of structure space, then
the simulation is restarted with the newly tuned potential. Here, we apply this
idea to the lattice configuration problem. 

Many forms of MLPs have been proposed in the literature; in this work, we employ
the high-dimensional neural network potential (NNP) scheme proposed by Behler
and Parinello \cite{behler2007,behler2011,behler2015,Watanabe2020}. This scheme assumes that
the total energy can be decomposed into atomic energies determined by the
environment around each atom and uses neural networks to fit these atomic
energies. That is, the total energy is expressed as
\begin{equation}\label{eq:NNP}
    E(\vec{\sigma}) = \sum_i^\text{atoms} \text{NNP}_{t_i} (f[\vec{\sigma}_i^{R_\text{c}}]),
\end{equation}
where the configuration vector $\vec{\sigma}$ is represented by continuous atom
coordinates in contrast to the cluster expansion (Eq.~\ref{eq:CE}) where
$\vec{\sigma}$ represents occupations on a discrete lattice. $t_i$ represents
the atom type, meaning that a unique NNP is fitted for each atom type.
$\vec{\sigma}_i^{R_\text{c}} \subset \vec{\sigma}$ represents coordinates of
atoms within a cutoff radius $R_\text{c}$ of atom $i$, and $f$ is a fingerprint function that
transforms $\vec{\sigma}_i^{R_\text{c}}$ into a multidimensional descriptor of
the atomic environment. To be physically meaningful and transferable, $f$ must
be chosen so that the resulting descriptor is invariant with respect to
translation and rotation of the system. To this end, several fingerprinting
schemes have been devised including {\it symmetry functions}
\cite{behler2007,behler2011,behler2015}, {\it smooth overlap of atomic positions
(SOAP)} \cite{Bartok2013}, and {\it Deep Potential} \cite{Zhang2018b,Zhang2018c}.
In this work, we employ the Chebychev
polynomial-based fingerprint proposed by Artrith {\it et al.}\cite{artrith2017};
the effectiveness of this fingerprint function has been
demonstrated especially for multicomponent systems. The effectiveness of other 
fingerprints or MLP forms for
our purposes is certainly of interest, and may be considered in future works.

The main distinguishing point of this work compared to the literature on
high-dimensional NNPs is that we train the NNP model to predict the total
energies of relaxed structures from unrelaxed ideal lattice structures with
configurational disorder. This is quite different from the usual approach of
learning the total energies as a function of continuous atom coordinates (the
only exception we are aware of is Ref.~\citenum{Li2017_Cu}, which used the NNP
scheme to describe the relaxation energy when a single Cu interstitial is
dissolved in amorphous Ta$_2$O$_5$). 
The obvious merit compared to conventional NNP approaches is that we can bypass
structural relaxation, while the demerit is that we lose access to ion
dynamics, and that we cannot expect transferability to other lattice structures.
Since relaxation calculations in a moderately sized supercell can
sometimes take hundreds of steps, the speedup attained in this manner can be
significant. On the other hand, the abovementioned demerits are not an issue when using
the NNP model for MC sampling on a lattice (it should be noted that MC sampling in
continuous coordinate space using random trial displacements is quite
inefficient because atoms are more or less close-packed and such trial steps will
almost never lead to exchange of atom sites \cite{Kasamatsu2019}). 

In a sense, this is a ``lazy'' alternative
to the cluster expansion method without the need to explicitly consider
optimized bases, i.e., clusters; there is also no need to perform any explicit
basis reduction based on lattice symmetry. This circumvents various issues associated
with choosing the finite number of clusters to be used for prediction that were mentioned
above.
Also, promising atom-environment descriptors have been proposed for treating
multicomponent and multi-sublattice systems without combinatorial explosion \cite{artrith2017},
which is an unsolved issue for cluster expansion.
We can also use the same
fingerprint functions for bulk, surface, and interface systems, while cluster expansion
requires more clusters for non-bulk systems \cite{Yuge2007}.
Another merit is that the nonlinearity of the NNP model can lead to better
convergence in practice compared to linear models (such as the cluster
expansion) \cite{artrith2017}. This may lead to more accurate description of 
systems with significant lattice relaxation where cluster expansion exhibits some
difficulties \cite{Nguyen2017}. We will explore this aspect in the following sections. 
A remaining issue is in the choice of input structures for
training, which is just as problematic as other approaches. In fact, we demonstrate a
rather spectacular failure when training only on randomly generated
configurations, and then show how well the active learning approach can solve
this issue.

As a side note, we point out that the NNP can be directly trained to predict the
energy as a function of the configuration vector $\vec{\sigma}$, as was done
in Ref.~\citenum{Ji2017}. However, the transferability of the NNP to larger
supercells in such a scheme is nontrivial, and thus we advocate the use
of atom-centric fingerprint functions. It is also possible to perform nonlinear
cluster expansion, i.e., fitting the total energy as a function of cluster
correlations using a neural network \cite{Natarajan2018}. Yet another successful
model is the low rank potential, which expresses the interatomic
potential as a low rank decomposition of a tensor whose indices
specify the local atomic environment \cite{Kostiuchenko2019,Shapeev2017}.
We do not claim that the approach presented in this work is an overall better method,
as it is difficult to compare overall performance including calculation speed,
accuracy, and ease-of-use on equal footing. However, as noted above,
cluster selection or explicit mapping to a tensor-like expression required in
other schemes can become intractable with 
increasing complexity of the system. Thus, we believe that the current approach
will enable configurational sampling in complex bulk or interface 
systems with many components and sublattices, i.e., those systems where it is
technically difficult to employ previously proposed approaches.

\section{Model and methods}
\subsection{Benchmark system: spinel oxides}
We demonstrate the above idea on the calculation of the degree of A/B-site inversion in
a few oxides with spinel-type structures. Spinel oxides share the general formula AB$_2$O$_4$
with A (B) representing a divalent (trivalent) cation, and are of interest in 
mineralogy as well as technological applications including magnets, photocatalysis, and battery materials
\cite{Biagioni2014,Thackeray2021,Narang2021}. In an ``ordered'' or ``normal'' spinel, all divalent cations
occupy the tetrahedrally coordinated sites, and all trivalent cations occupy the octahedral sites.
However, it is known that spinels with various divalent/trivalent cation combinations can exhibit
significant deviation from the ``normal'' occupation, and this is often quantified by the degree of inversion (DOI) defined as
the fraction of A-sites (tetrahedral) occupied by B (trivalent) cations. The DOI corresponding to 
a completely random occupation is $2/3$, while a nearly ``inverse'' occupation with 
DOI of nearly unity is also known for some compounds.
Prediction of the DOI as a function of temperature is of interest because of its potential 
impact on the magnetic, optical, thermal, and
electrochemical properties of the material.

In this work, we calculate the degree of inversion in
MgAl$_2$O$_4$, ZnAl$_2$O$_4$, and MgGa$_2$O$_4$ and compare to a series
of cluster expansion works by Seko {\it et al.} \cite{Seko2006,Seko2009,Seko2011,Seko2014}
The former two are known to be nearly normal, while significant inversion has been
reported for MgGa$_2$O$_4$.
Spinels constitute a prototypical multivalent system (i.e.,
A$^{2+}$ and B$^{3+}$ share lattice sites) where a naive cluster expansion can
lead to large qualitative errors due to overfitting in a small calculation cell;
augmentation with a screened point charge model (CE-SPCM) was found to be
necessary to obtain accurate predictions without considering rather long-range
pair clusters in Ref.~\onlinecite{Seko2014}. Also, the feasibility of avoiding
cluster expansion fitting and sampling directly on DFT-relaxed energies was
demonstrated in combination with RXMC sampling in
Ref.~\onlinecite{Kasamatsu2019} with a 48-cation model. Here, we consider a
192-cation model, which is completely beyond the reach of direct DFT sampling
for two reasons: (1) because of the longer time required for each DFT
calculation, and (2) the {\it much} larger configurational space that has to be
sampled. Regarding (2), the number of degrees of freedom for the 48 cation model
is $_{48}\text{C}_{16} \sim 2 \times 10^{12}$, while that for the 192 cation model
is $_{192}\text{C}_{64} \sim 7 \times 10^{51}$. This explosion in the configuration space
should attest to the curse of dimensionality that we are fighting against,
although there is actually no need to perform calculations on all of these 
configurations if good sampling schemes (such as RXMC) are available.

\subsection{DFT calculations}
The reference DFT calculations were performed using VASP code \cite{Kresse1996,Kresse1996a}. We employed the projector
augmented wave method \cite{Blochl1994} to describe electron-ion interactions. A plane wave basis
set with an energy cutoff of 400 eV was used to expand the wave functions. The
Brillouin zone was sampled only at the $\Gamma$ point. The GGA-PBE functional \cite{Perdew1996}
was used to approximate the exchange-correlation energy. In this work, we performed relaxation of the 
lattice vectors as well as the internal coordinates to within 0.04 eV/\r{A}. It should be noted that 
lattice vector relaxation leads to changes in the calculation mesh density, which means that the effective cutoff 
energy deviates from the preset value. To alleviate this issue, the relaxations were 
repeated several times so that the final structures and energies are consistent with the cutoff energy of 
400 eV.

\subsection{NNP training and evaluation}\label{Sec:NNPdetails}
The NNP training and subsequent evaluation were performed using Atomic Energy NETwork ({\ae}net) code
\cite{artrith2016, artrith2017}. As noted above, the NNP model
is trained to predict relaxed energies. This may be represented using a slightly
modified version of Eq.~\ref{eq:NNP}:
\begin{equation} \label{eq:latNNP}
    E_\text{rel}(\vec{\sigma}) = \sum_i^\text{atoms} \text{NNP}_{t_i}^\text{rel} (f[\vec{\sigma}_i^{R_\text{c}}]) 
    \text{ for } \vec{\sigma} \in \{\vec{\sigma}_\text{lattice}\},
\end{equation}
where $E_\text{rel}(\vec{\sigma})$ represents the energy after structural
relaxation from a starting structure $\vec{\sigma}$, 
$f[\vec{\sigma}_i^{R_\text{c}}]$ represents atomic environment descriptors for
the starting structure, and $\text{NNP}_{t_i}^\text{rel}$ represents atomic
energy contributions to the relaxed total energy. The training and use of the
NNP is restricted to $\vec{\sigma} \in \{\vec{\sigma}_\text{lattice}\}$, where
$\{\vec{\sigma}_\text{lattice}\}$ represents the set of ideal lattice structures
with configurational disorder.
Equation \ref{eq:latNNP} states that the relaxed energies can be
calculated as a sum of atomic contributions, which, in turn, can be calculated
from coordinates corresponding to the ideal lattice. 
This is an ansatz that we are making. In general, it
is impossible to prove the exactness of a certain machine learning model in
reproducing a physics model, and our approach is no exception. This is in
contrast to cluster expansion for which a rigorous proof exists
for mapping fully relaxed energies to a lattice-based Hamiltonian
\cite{Ceder1993,Sanchez2017}.
The ansatz will need to be verified for each system under study, and active
learning is an ideal approach in this regard.

We used the fingerprinting scheme by Artrith {\it
et al.} \cite{artrith2017}, where the radial and angular distribution functions
(RDF and ADF) with an appropriately chosen cutoff are expanded by an
orthogonal basis set based on Chebychev polynomials, and the expansion
coefficients are fed in to the neural network as the input descriptors. 
A key point in their scheme is that the structural and compositional descriptors
are given separately; the structural descriptor is given by the usual RDF and
ADF, and the compositional descriptor is also given by RDF and ADF but with
atomic contributions weighted differently for each chemical species. In this
work, we expand the RDF with a cutoff of 8.0 \r{A} and
expansion order 16 and the ADF with a cutoff of 6.5 \r{A} and
expansion order 4. The employed neural network thus has an
input layer with 44 nodes\footnote{We have 16 coefficients for RDF and 4 for
ADF resulting in 20 descriptors, and this is multiplied by 2 since we have separate
RDFs/ADFs for composition and structure.}, and we chose to use 2 hidden layers with
15 nodes each and the $\tanh$ activation (there is actually an additional bias
node to accommodate arbitrary references for the energy \cite{artrith2016}). 
The training epocs were monitored and
terminated when the test set error began to increase. We note that it is 
necessary to validate the model on an additional data set that is separate from 
testing data used for stopping of the training epocs. 
This is done, in a sense, through the active learning procedure
detailed below.

\begin{figure}
    \centering
    \includegraphics[width=\columnwidth]{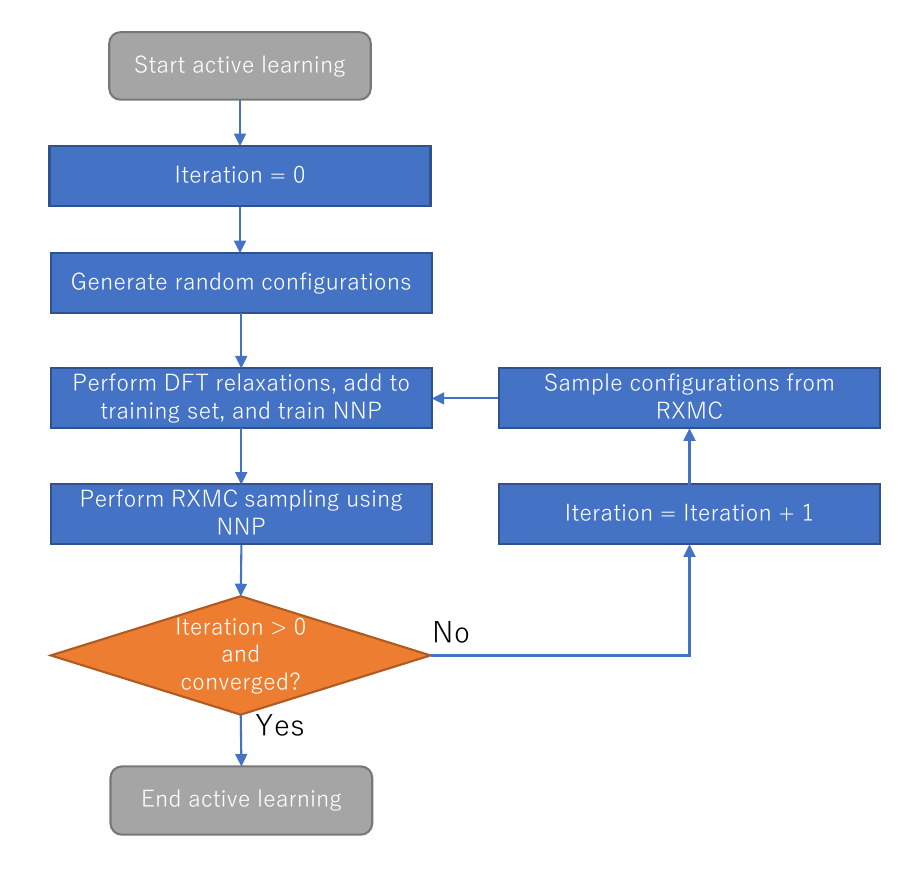}
    \caption{A flowchart of our configurational sampling scheme using active learning. 
    }
    \label{fig:scheme}
\end{figure}
\subsection{Replica exchange Monte Carlo sampling combined with active learning} \label{sec:AL}
The configurational sampling was performed using the replica exchange Monte Carlo 
(RXMC) method combined directly
with NNP evaluation using {\ae}net. 
The RXMC method employs several copies, or replicas of the system under study.
Each of the replicas are sampled using the usual Metropolis MC algorithm at
different temperatures. At preset intervals, temperatures are swapped between
replicas according to the Metropolis criterion with the following probabilities:
\begin{equation}
    P = \min \{1, \exp[(\beta_i - \beta_j) (E_i - E_j)] \}.
\end{equation}
Here, $i, j$ are the replica indices, $\beta$ is the inverse temperature
$1/(k_\text{B}T)$, and $E$ is the energy of the replica. Higher temperature
replicas are given the task of global surveillance of the energy landscape while
lower temperature replicas basically perform local optimization. The swapping of
temperatures occurs when a higher temperature replica finds a new energy minimum;
local optimization is then performed at the lower temperature.
The detailed theory behind RXMC is given in Ref.~\citenum{Hukushima1996}. Our
previous works \cite{Kasamatsu2019,Kasamatsu2020} may also be helpful for
understanding the basic concept.
A parallelized code for RXMC calculations in combination with DFT codes
as well as {\ae}net is available as a part of ab-Initio Configuration
Sampling Toolkit (abICS) \url{https://github.com/issp-center-dev/abICS}. 
In this work, 15 replicas of the system were sampled in parallel with temperatures ranging from
400 K to 1900 K, and temperature exchange was attempted every 4 steps to speed
up the global sampling. Equilibration was performed for 
80000 steps, then the DOI, which is calculated as the ratio of Al ions on Mg 
sites, was averaged from 400000 steps to obtain the temperature dependence. 

The NNPs used in the RXMC sampling were refined iteratively using the active
learning scheme shown in
Fig~\ref{fig:scheme}. We note that a very similar process using an
effective pair interaction model was reported recently
\cite{Liu2021}.
First, we train an NNP set with randomly generated samples, then use
those NNPs to perform RXMC sampling. In subsequent iterations,
we sample a preset number of configurations from the previous RXMC run and
perform DFT relaxations on those configurations. Comparison of the DFT energies with the
NNP predictions serves as the validation procedure for the machine-learning
model, although the validation
data set will be biased somewhat as we discuss below. Discrepancies
between DFT energies and NNP predictions will signify that the simulation
has wandered into previously unlearned regions of configuration space. 
We then add those DFT results to the training set, retrain the NNPs, and run
RXMC sampling again. This process is repeated until convergence is achieved
in the physical quantities of interest.
In this work, we sampled 300 structures from the first 6400 RXMC steps in 
each active learning iteration. The final DOI's were calculated from a separate
longer RXMC run with 480000 RXMC steps (80000 for equilibration and 400000 for 
thermodynamic averaging) as mentioned above.

We note that the above procedure biases the training data set towards
thermodynamically relevant configurations predicted by the NNP model in that active learning step.
The thermodynamic biasing, in itself, is usually not a bad thing since 
we are often only concerned with thermodynamically relevant, i.e., realistic
configurations. However, the resulting NNPs cannot be expected to perform well
when used to predict energies of configurations with both low entropy and high energy, 
since such configurations are thermodynamically unstable over all temperatures and
rarely visited in the RXMC procedure. Such an example in the case of AB$_2$O$_4$
spinels considered in this work would be a phase-separated configuration with ``A'' cations in 
one half of the cell and ``B'' cations in the other half. 
If, for some reason, we want to correctly reproduce the energy of such situations, we would have to 
add such configurations to the dataset, or we may accumulate training data using active learning with
varying compositions.

On the other hand, the fact that we are relying on an inaccurate NNP model to generate the
training set during the active learning iterations may pose issues with uncontrolled biasing of the
training data.
In the case where the model underestimates the energy of certain configurations, the RXMC algorithm at
lower temperatures will
bias the sampling towards those configurations so that the underestimation is quickly corrected in 
the subsequent iteration. The biggest foreseeable problem is when the model overestimates the energy.
This is covered to some extent by setting a high enough temperature for the RXMC sampling so that 
configurations with higher predicted energies will still be sampled. It is also noted that RXMC in each iteration
is initialized with random configurations, i.e., with an effective temperature of infinity. Ultimately,
it is necessary to evaluate if such issues are severe through computational experiments, which we present in 
the following sections.

We should also mention that we are using the term ``active learning'' in a rather broad sense,
that the training set configurations are not given {\it a priori} but are chosen according to
the given calculation procedure.
In our case, this is done through thermodynamic importance sampling as explained above.
On the other hand, many works in the active learning MLP literature employ various uncertainty estimates 
to determine if a structure met during simulation is in an extrapolation region and should 
be added to the training set \cite{jinnouchi2020,Podryabinkin2017}. 
The efficacy of using such uncertainty indicators to 
further limit the number of expensive {\it ab initio} calculations 
may be explored in the near future.

\section{Results and discussion}
\begin{figure*}
    \centering
    \includegraphics[width=2\columnwidth]{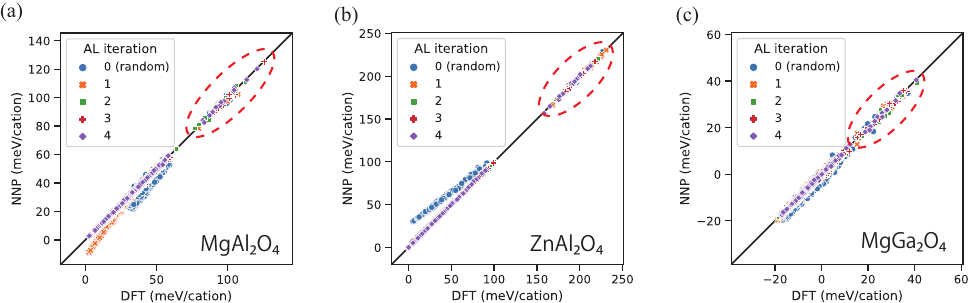}
    \caption{Improvement in NNP predictions vs.~DFT reference energies with the number of 
    active learning iterations in the three spinel oxides, (a) MgAl$_2$O$_4$, (b) ZnAl$_2$O$_4$,
    and (c) MgGa$_2$O$_4$. RXMC runs are performed using the NNP, then DFT calculations are performed on 
    structures from the same RXMC run to evaluate the prediction accuracy.
    The zero reference for the energy is taken to be the DFT energy for the ordered spinel 
    configuration. Data on the diagonal line correspond to high prediction accuracy. The dashed
    circles indicate energy regions of randomly generated configurations in the first steps of each 
    active learning RXMC iteration \cite{Note2}.}
    \label{fig:NNP_AL_ecorr}
\end{figure*}

\begin{figure*}
    \centering
    \includegraphics[width=2\columnwidth]{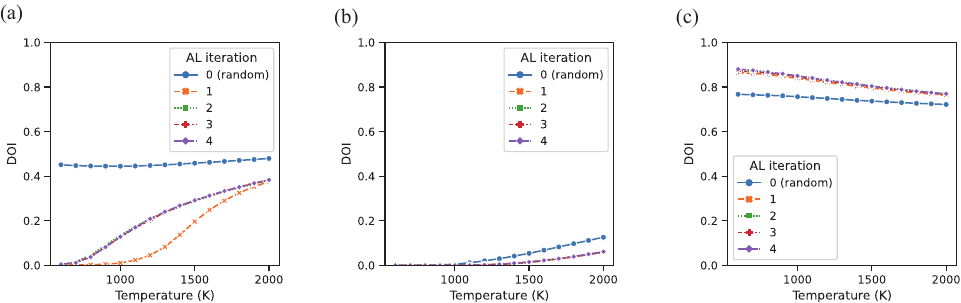}
    \caption{The convergence of the calculated temperature-dependence of the degree
    of inversion vs.~active learning iterations in 
    the three spinel oxides, (a) MgAl$_2$O$_4$, (b) ZnAl$_2$O$_4$,
    and (c) MgGa$_2$O$_4$. 
    \label{fig:DOI_conv}}
\end{figure*}

\subsection{Active learning}
We started the active learning process with 300 randomly chosen configurations.
DFT relaxation and energy calculations were performed on these configurations,
then an NNP model was trained to predict relaxed energies from the input
configurations. The oxygen sublattice is neglected in the training process since
the input structures all have identical oxygen coordinates and do not provide
any useful information for the prediction. We used 90\% of input structures for
training and 10\% for testing. 
For the spinel systems studied in this work, we typically obtained mean 
absolute errors (MAEs) and root mean squared errors (RMSEs) of less than 
a few meV/cation 
in the training and test sets, which is close to the accuracy
limitations of the DFT calculations themselves.
However, such low fitting error does not guarantee high accuracy in predicting the energies
of structures obtained in the RXMC sampling. This is checked by performing DFT relaxations
on the structures encountered during RXMC sampling using the NNP.
As shown in Fig.~\ref{fig:NNP_AL_ecorr}, 
the NNP trained on random samples (indicated by blue circles in Fig.~2) shows poor
accuracy, especially in predicting the energies of the lower energy configurations.
This is because
the model was trained on random configurations with higher energies, 
and the training set did not include lower energy structures with 
less disorder 
(please note that the splitting of samples between higher and
lower reference DFT energies seen here and in subsequent figures originates from our sampling
procedure and is not an inherent property of the system.\footnote{The splitting
of samples between higher and lower energies with few samples
in the interval of e.g., $\sim$ 50--75 meV/cation for MgAl$_2$O$_4$ originates from our sampling
procedure: we always sampled the first step in the RXMC calculations which is
generated randomly, and most of the replicas relax quickly to lower energy
configurations before the next sampling step.}). 
Increasing the number of random samples does not improve the results as we show later. 
The importance of having configurations with varying degrees of order in the training set
has already been pointed out in the cluster expansion literature. A
suggested solution was to perform a grouping of input structures based on
correlation functions (which roughly corresponds to the degree of disorder), and
to perform cross validation on each group during the fitting process \cite{Seko2011}. 
Here, we take an
alternate approach, i.e., active learning, as was detailed in Sec.~\ref{sec:AL}.
The predictions at lower energies progressively improve with the 
number of active learning iterations because of the thermodynamic biasing mentioned above; the prediction RMSE for the three 
spinel systems studied here converged within 0.5 meV after the third iteration.
This is clearly smaller than $\sim$10 meV/cation 
reported in Ref.~\citenum{Seko2014} using cluster expansion, although it may not
be a fair comparison because of the difference in the data set used to calculate
prediction errors. 

%Also, the better performance may be coming more from the 
%larger training set with larger supercells made possible by today's supercomputing 
%resources rather than the methodology itself.

\begin{figure}
    \centering
    \includegraphics[width=\columnwidth]{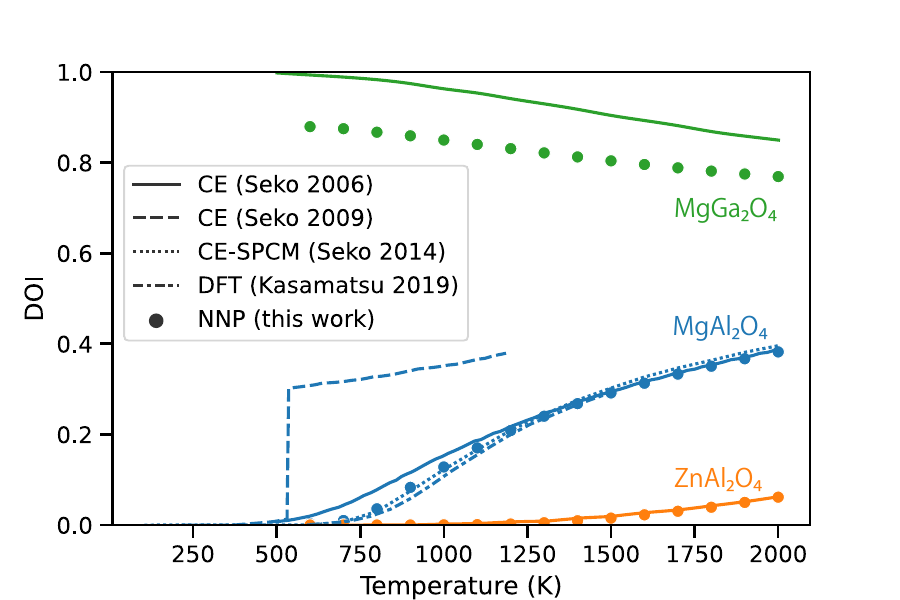}
    \caption{The degree of inversion calculated in the present work for the 
    three spinel oxides MgGa$_2$O$_4$ (green), MgAl$_2$O$_4$ (blue), and ZnAl$_2$O$_4$ (orange)
    in the 192-cation cell whose neural network model was trained in the 192-cation cell 
    ($\bullet$). The results are also compared to the 
    cluster expansion (CE) literature \cite{Seko2006,Seko2009,Seko2014} and our previous
    work using DFT calculations directly in combination with RXMC \cite{Kasamatsu2019} 
    in a 48-cation model (lines).
    }
    \label{fig:DOI}
\end{figure}

Convergence is also attained in the physical quantity of interest, i.e., the 
DOI, as shown in Fig.~\ref{fig:DOI_conv}. For the three spinel oxides
treated here, 2 or 3 active learning iterations with 300 DFT relaxation calculations 
each were sufficient to obtain a converged DOI plot (Fig.~\ref{fig:DOI_conv}, purple 
diamonds), which looks
completely different from that predicted using NNPs trained on randomly generated
configurations (Fig.~\ref{fig:DOI_conv}, blue circles).
The final DOI plots are in good agreement with most of the cluster expansion literature
as well our previous work employing direct RXMC sampling on DFT energies (Fig.~\ref{fig:DOI}).
There is, however, a sizeable deviation for MgGa$_2$O$_4$, and there is one cluster expansion work
for MgAl$_2$O$_4$ that deviates from the others and shows a discontinuous transition around
600 K \cite{Seko2009} (denoted by ``Seko2009'' in Fig.~\ref{fig:DOI}). 
As for MgGa$_2$O$_4$, the current work predicts a DOI which
is about 0.1 smaller than cluster expansion in Ref.~\onlinecite{Seko2006}. 
Our result is within the range of most of the available experimental data that report 
0.75 \cite{Machatschki1932}, 0.67 \cite{Huber1957}, 0.84--0.90 \cite{Schmalzried1961},
and 0.81--0.84 \cite{Weidenborner1966}, although the first work that put forth the idea of an 
``inverse spinel'' reported unity, i.e.,
complete inversion \cite{Barth1932}. 
The discrepancy may come from two sources apart from the difference in the employed models
(i.e., cluster expansion and NNP): the supercell size used in the training stage, and that used in
the MC sampling stage. The cluster expansion literature has performed ECI fitting in supercells with
24 cation sites or smaller, while they performed MC simulations on supercells with as many as 
48000 cations. Here, both the training and sampling were performed on 192-cation supercells.
The impact of these supercell sizes on the results are discussed in more detail in 
Sec.~\ref{sec:cellsize}. 
As for ``Seko2009'' \cite{Seko2009},
the discontinuous transition was
reported in a later work to be a spurious result due to truncation of 
the range of pair clusters in the cross validation process \cite{Seko2014}.

\begin{figure*}
    \centering
    \includegraphics[width=1.3\columnwidth]{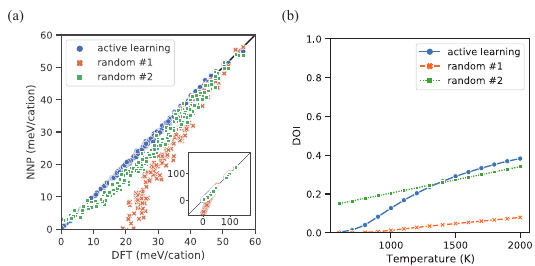}
    \caption{(a) Comparison of NNP predictions vs. DFT reference energies when the NNP was 
    trained on 1200 configurations generated by the active learning procedure (blue circles) or
    random sampling (green squares and orange crosses corresponding to two independent
    training runs on the same randomly generated dataset). The comparison is performed 
    on the dataset taken from the third active learning iteration 
    (same data as Fig.~2a, red $+$). The inset shows all data points, while the main figure
    shows a magnified version for clarity. (b) The degree of inversion calculated for 
    MgAl$_2$O$_4$ using the three NNPs compared in (a).
    }
    \label{fig:DOIrand}
\end{figure*}

To confirm that the active learning procedure is indeed more efficient than random
sampling, we performed DFT relaxation on 1200 randomly generated configurations of MgAl$_2$O$_4$,
trained the NNP on those configurations, and compared the results to those from active
learning with the same number of DFT calculations (1 random + 3 active learning iterations, which was
enough to converge the DOI results as shown in Fig.~\ref{fig:DOI_conv}a).
The training on the random dataset was performed independently twice 
with different training/test data separation and initial neural network weights. 
As shown in Fig.~\ref{fig:DOIrand}a,
the energy predictions from NNPs trained on random configurations deviate from the DFT value
for configurations with lower energies, and it is also clearly less accurate
than the NNP trained through the active learning procedure. Again, this is not surprising because
lower energy configurations are seldom included in the randomly generated training set.
Moreover, the deviation behavior is rather uncontrolled as indicated
by different predictions in the energies and the DOIs (Fig.~\ref{fig:DOIrand}b) 
for the two independent training runs; thus, it is virtually impossible to obtain the
correct temperature-dependent results with only random generation of the training dataset.

Savings in computer time achieved by the current approach turned out to be
quite significant compared to sampling directly on DFT energies. 
On the Xeon E5-2680v3 CPU, the NNP prediction using 1
CPU core is a few hundred to a few thousand times faster than DFT relaxation
using 216 cores depending on the number relaxation steps necessary for convergence. 
We also note that the number of necessary DFT calculations turned out
to be orders of magnitude smaller than the number of NNP evaluations, i.e., RXMC steps,
necessary for converging the
DOI. A converged DOI was obtained for ZnAl$_2$O$_4$ and MgGa$_2$O$_4$ at the first active learning iteration,
while the DOI for MgAl$_2$O$_4$ was converged at the third active learning iteration (Fig.~\ref{fig:DOI_conv}), 
which correspond to 600 and 900 DFT relaxations, respectively. On the other hand,
$\sim7\ 200\ 000$ NNP evaluations ($=480\ 000 \text{ MC
steps/replica} \times 15 \text{ replicas}$) were necessary to obtain a converged
DOI vs.~temperature. The $\sim 1000$ DFT calculations can be completed within a few
hours on a modern supercomputer system since the calculations can be performed
completely in parallel. The RXMC calculations took roughly one day
using 15 CPU 
cores, which would fit in one node of a modern workstation. 
We also note that the RXMC
simulations were initialized with random configurations, and yet they still
found the single ground state ordered spinel configuration out of $\sim10^{51}$
possible configurations for the normal spinels MgAl$_2$O$_4$ and ZnAl$_2$O$_4$.
This attests to the effectiveness of modern sampling
methods in battling the curse of dimensionality.

The fact that less than 1000 reference samples were necessary to obtain
an accurate NNP model might be surprising, as previous works on NNPs usually
report at least one order of magnitude larger reference sets (see, e.g.,
Ref.~\cite{artrith2016,Zhang2018c}). This is most likely because the input
configuration space is restricted to $\vec{\sigma} \in
\{\vec{\sigma}_\text{lattice}\}$, and this is a clear merit of
our approach. It should be pointed out, however, that each of our reference
samples is a result of DFT relaxation calculations, which comes at an additional
cost compared to conventional NNP approaches that can be trained on structures without
relaxation. Still, it should be {\it much} easier to attain sufficient sample
coverage in this restricted configuration space. The restriction also makes the
problem easier for the neural network, as it needs to distinguish between {\it much}
fewer structures. These facts combined with the ability to bypass 
lattice relaxation make the present approach more suitable for the lattice
configuration problem in comparison to conventional continuous-coordinate NNP methods. 
That is not to say that those methods cannot be applied to systems with
configurational disorder. In fact, the {\it Deep Potential} approach has been
applied to randomly generated high entropy alloys \cite{Zhang2018c, Dai2020}
with promising results, and it can provide vibration properties
while the present approach cannot. However, the computational expense for NNP
training and simulation of configurational order/disorder behavior will be much
larger than our lattice-restricted approach, and they are yet to be applied to
this type of calculation.

\begin{figure*}
    \centering
    \includegraphics[width=2\columnwidth]{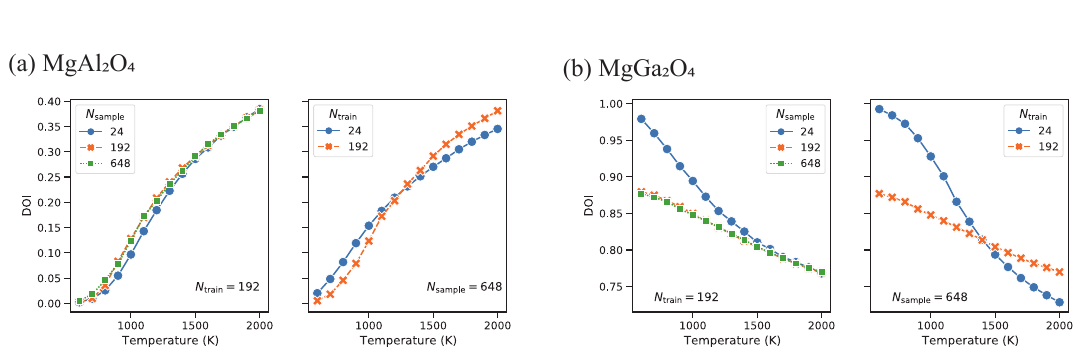}
    \caption{The dependence of the calculated degree of inversions (DOIs) in (a) MgAl$_2$O$_4$
    and (b) MgGa$_2$O$_4$ vs.~training cell size (denoted by the number of cations in the training cell $N_\text{train}$) 
    and MC sampling cell size (denoted by the number of cations in the sampling cell $N_\text{sample}$).}
    \label{fig:train_sample_DOIs}
\end{figure*}

\subsection{Convergence vs.~supercell size used for training and Monte Carlo sampling}
\label{sec:cellsize}

The usual approach in the NNP and cluster expansion literature is to use a set of
many small-scale DFT calculations to fit the model and use that model to accelerate 
calculations in an expanded supercell. In doing so, one must be careful that (1) the
fitting procedure does not lead to overfitting in the small cells at the cost of
transferability to larger cells not contained in the training set, and (2)
thermodynamically relevant correlations are contained within the training set supercell 
size. These points are especially important here since spinel oxides are mulitvalent
ionic oxides where significant long-range interactions are expected.
% An additional point that we were interested in is how well the NNP trained on a
% smaller supercell can describe configuration statistics in larger supercells, as
% we were worried about overfitting to structures in a small cell. 

In the case of cluster expansion,
cross validation is usually performed within the training set to truncate
the number of clusters to a computationally tractable amount, but for systems 
such as MgAl$_2$O$_4$,
this procedure can lead to large prediction errors for structures with 
longer periods than those in the training set \cite{Seko2014}. 
This lead to a spurious prediction of 
a discontinuous order-disorder transition in MgAl$_2$O$_4$
at $\sim$600 K in Ref.~\onlinecite{Seko2009} 
(denoted by ``Seko 2009'' in Fig.~\ref{fig:DOI}); it was found to be
necessary to either ignore the truncation procedure and use as many
long-distance pair clusters as possible, or to use fewer pair clusters and 
augment the cluster expansion Hamiltonian with a screened Coulomb term, which lead to 
prediction of a more continuous transition in Ref.~\onlinecite{Seko2014} (``Seko 2014''
in Fig.~\ref{fig:DOI}). 
It is difficult to see if similar issues may arise in our approach. Although
there is no explicit truncation procedure, neural networks are prone to 
overfitting to the available data; the network might decide that interactions beyond
the training set supercell dimensions are unimportant and effectively truncate
the range of interactions. Even if it
does not, we would have to see if the cutoff radii of 8 \r{A} for the radial
descriptor and 6.5 \r{A} for the angular descriptor was long enough for
sufficient precision. 

\begin{figure}
    \centering
    \includegraphics[width=\columnwidth]{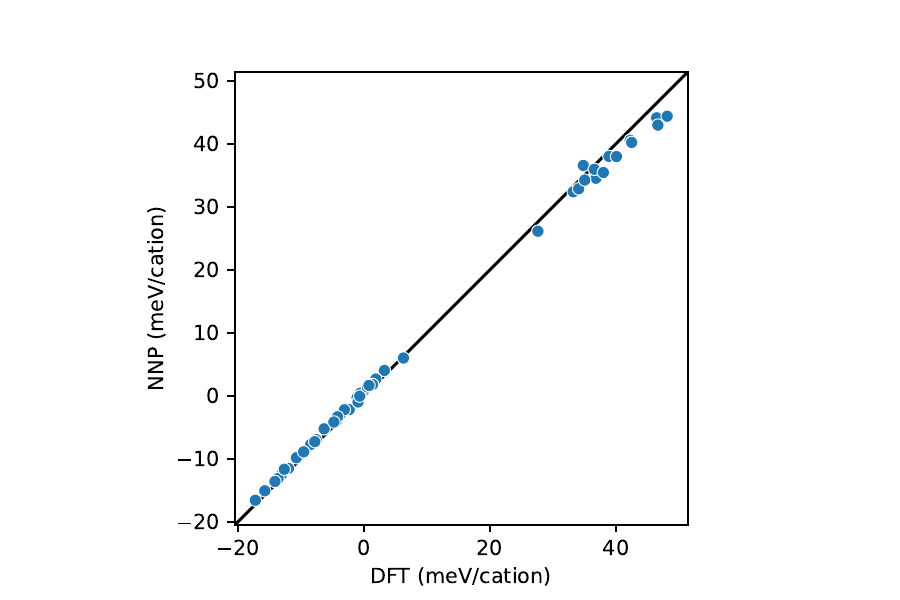}
    \caption{Correspondence between NNP prediction and DFT results for 648-cation cell MgGa$_2$O$_4$
    configurations when the NNP was trained in the 192-cation cell.}
    \label{fig:MGOx3ref}
\end{figure}

To check for such issues, we examined the convergence behavior against the supercell 
size used for training and
MC sampling as follows. First, we performed active learning of the NNP model in a conventional cell 
of cubic spinel having 24 cation sites in addition to the 192-cation model examined above.
Two active learning iterations were performed to converge the results for the 24-cation model.
Then, we performed 
RXMC sampling calculations on the 24-cation, 192-cation, and 648-cation models using those NNP models.
This means that we performed 6 RXMC calculations resulting in DOI data 
that can be indexed by the number of cation sites 
in the training and sampling steps as
$\text{DOI}(N_\text{train}, N_\text{sample})$.
Figure \ref{fig:train_sample_DOIs} shows the results for MgAl$_2$O$_4$ and MgGa$_2$O$_4$, where
the left hand side plot shows the dependence vs.~$N_\text{sample}$ at $N_\text{train}=192$ and the 
right hand side plot shows that against $N_\text{train}$ with $N_\text{sample}=648$.
It is clear that we have attained convergence against RXMC sampling cell size, as the results for
$N_\text{sample}=192$ and 648 are virtually identical, and $N_\text{sample}=24$ also results in 
a decent prediction in the case of MgAl$_2$O$_4$.
This is surprisingly small compared to the supercell sizes employed for MC calculations in
the cluster expansion literature (48000 as mentioned above), but we have obtained essentially 
identical results for MgAl$_2$O$_4$ and ZnAl$_2$O$_4$ as seen in Fig.~\ref{fig:DOI}.
On the other hand, the convergence with respect to the training supercell size $N_\text{train}$ 
is more difficult to confirm.
Sizeable differences are seen for $N_\text{train}=24$ and 192, 
especially for MgGa$_2$O$_4$. This shows that the NNP trained
in the 24-cation supercell of MgGa$_2$O$_4$ is not capable of reproducing the 
correct energetics in larger supercells.
Now, the ideal way to check whether we have achieved convergence at $N_\text{train}=192$
would be to perform active learning on 
a larger supercell, but that would be prohibitively expensive in terms
of computer resources. As a compromise, we decided to test the accuracy of the $N_\text{train}=192$ NNP model 
by performing DFT relaxations on a small subset of the 648-cation configurations from 
the $(N_\text{train}, N_\text{sample})=(192,648)$ calculation. As shown in Fig.~\ref{fig:MGOx3ref}, the $N_\text{train}=192$ NNP
turns out to be quite accurate in predicting the energies in the 648-cation cell. Thus, we can be 
fairly confident that the results are converged with respect to the training set supercell
size at 192 cations. The discrepancy of the MgGa$_2$O$_4$ DOI with Ref.~\onlinecite{Seko2006} is
at least partly due to the fact that they employed only up to 24-cation supercells
in the fitting of the ECIs.

These results show that regardless of the model used (NNP or cluster expansion),
the training set needs to include structures with periods that are long
\emph{enough} for the system under study. Also, if the model includes hyperparameters
pertaining to the range of interactions, then that must be set long enough as well.
For spinel oxide compounds
examined here, 
the periods were found to be rather long but still tractable with the help of modern-day
supercomputers.
This may be surprising because our NNP model (or cluster expansion, for that matter) does not consider
truly long-range Coulomb interactions along the lines of, 
e.g., Ref.~\onlinecite{Artrith2011}, as all
explicit interactions are truncated by the cutoff in the 
descriptors. However, our results suggest that the cutoff is long enough, especially for
lower energy or higher entropy structures encountered during the Monte Carlo procedure which
should have relatively small polarization. The structural relaxation should also
reduce the required cutoff due to dielectric screening.

% This suggests that our NNP approach combined with the
% Chebychev fingerprint with moderate cutoff radii is doing a fairly good job of
% balancing short to long-range interactions, even when trained in a relatively
% small unit cell. That is not to say that our approach considers truly long-range
% Coulomb interactions along the lines of, e.g., Ref.~\onlinecite{Artrith2011}, as all
% explicit interactions are truncated by the cutoff in the 
% descriptors. However, our results suggest that the cutoff is long enough, especially for
% lower energy or higher entropy structures encountered during the Monte Carlo procedure which
% should have relatively small polarization. The structural relaxation should also
% reduce the required cutoff due to dielectric screening.
\begin{figure}[t]
    \centering
    \includegraphics[width=\columnwidth]{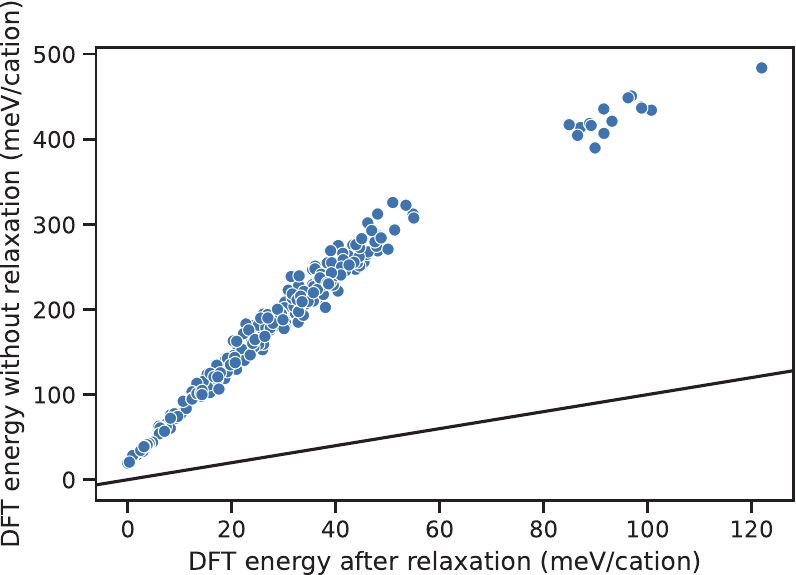}
    \caption{Energies before relaxation vs.~fully relaxed energies for
    samples taken from the active learning procedure on MgAl$_2$O$_4$. The solid
    line is a guide for the eye and corresponds to zero relaxation energy. 
    }
    \label{fig:relnorel}
\end{figure}

\subsection{Robustness against relaxation}

\begin{figure*}
    \centering
    \includegraphics[width=2\columnwidth]{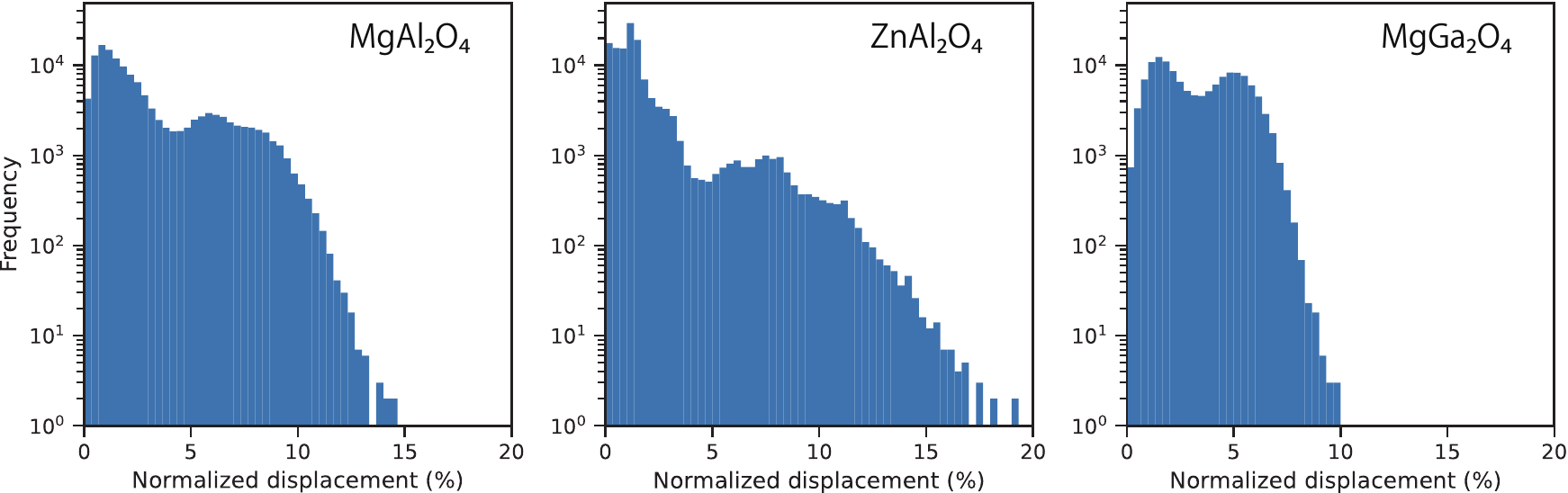}
    \caption{Histograms of normalized displacements of atoms upon relaxation in the spinel oxides under study.
    The displacements are calculated by performing DFT relaxation on 300 structures from the fourth (final) active learning iteration, and
    the histograms are generated from all atoms in those structures.
    %The arrows indicate the maximum displacement in the dataset for each system.
    }
    \label{fig:displacements}
\end{figure*}

\begin{figure}
    \centering
    \includegraphics[width=0.9\columnwidth]{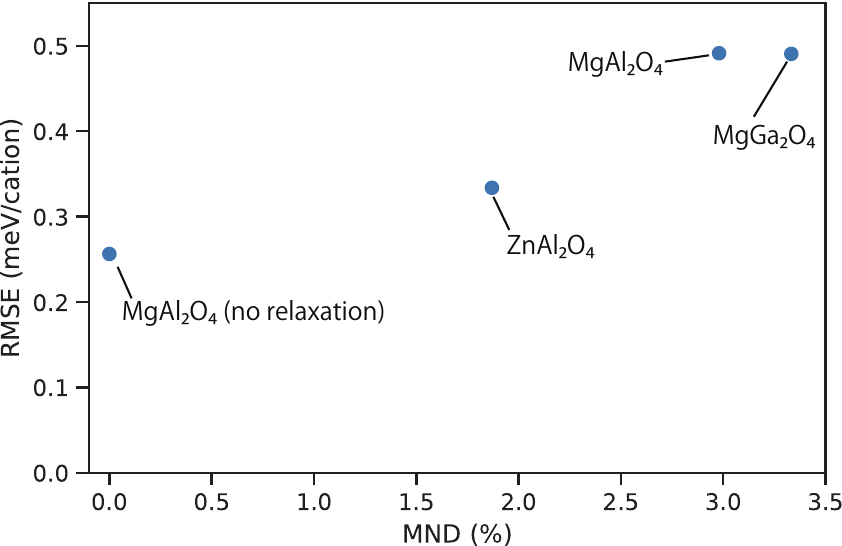}
    \caption{Prediction RMSEs vs.~mean normalized displacements due to relaxation in the spinel
    oxides calculated for the final active learning iteration.
    }
    \label{fig:relaxerror}
\end{figure}

One point worth stressing is that the success of the present approach
in predicting the relaxed energies from ideal lattice structures
is not a trivial consequence of negligibly small relaxation energies in this system.
As
shown in Fig.~\ref{fig:relnorel} for MgAl$_2$O$_4$, 
the relaxation energies are roughly one order of
magnitude larger than energy differences between relaxed configurations (notice
the difference in scales of the horizontal and vertical axes). Also, although the relaxation
energies are correlated with energies before relaxation, the correspondence is
not one-to-one and the scatter is as large as 50 meV/cation. This is clearly
significant when comparing against relaxed energy differences between distinct
configurations.

We also examine the magnitude of relaxation, as it has been shown to be correlated 
with prediction errors in the
cluster expansion literature \cite{Nguyen2017}.
To quantify the amount of relaxation in the three spinel systems under equal footing, 
we use the normalized displacement, which we define as
\begin{equation}
    \text{ND} = \frac{\Delta d}{V_0^{1/3}},
\end{equation}
where $\Delta d$ is the atom displacement upon relaxation and $V_0$ is the volume per atom \emph{after}
volume relaxation;
$V_0^{1/3}$ can be interpreted as an average lattice parameter, so the normalized displacement
defined in this manner is an indicator of local strain. As shown in Fig.~\ref{fig:displacements}, 
the relaxation in the three spinel oxides are far from negligible. A significant number of 
atoms show ND's larger than 5\%, and the maximum ND reaches  $\sim$19\% for ZnAl$_2$O$_4$.
The correlation between relaxation and prediction errors are shown in 
Fig.~\ref{fig:relaxerror}, where the prediction RMSEs for the three systems are 
plotted vs.~mean normalized displacement (MND),
which is calculated by averaging the normalized displacements over all atoms from all 
configurations in the dataset (Fig.~\ref{fig:displacements}). 
Active learning calculations without relaxation 
were also performed on MgAl$_2$O$_4$ to examine the prediction errors not attributed to relaxation. 
As a result, we see an increasing trend in the errors vs.~MND,
suggesting that
relaxation does lead to higher errors. However, the RMSEs are all less than
0.5 meV per atom, which is as good as we can reasonably expect when
considering numerical noise in the
relaxation process. 

We also note that the normalized mean \emph{squared} displacement (NMSD)
was suggested in Ref.~\onlinecite{Nguyen2017} as a relaxation metric
that correlates somewhat with prediction errors, and they heuristically categorized NMSD
values into ``Good'', ``Maybe'', and ``Bad'' regions in terms of expected
reliablity and quality of cluster expansion fitting. The NMSDs for MgAl$_2$O$_4$,
ZnAl$_2$O$_4$, and MgGa$_2$O$_4$ calculated from the active learning procedure are
0.158\%, 0.156\%, and 0.147\%, respectively, and they would all fall into the ``Maybe''
category. The fact that we were able to obtain RMSEs of less than 0.5 meV/atom 
for those systems indicates that
our neural network scheme is at least as robust as
cluster expansion against relaxation.
Ultimately, we would need to accumulate experience in applying this approach to 
a wider variety of systems, but that is beyond the scope of this work.

%\subsection{Robustness against lattice relaxation}

\subsection{Robustness with respect to input precision}

\begin{figure}[ht]
    \centering
    \includegraphics[width=\columnwidth]{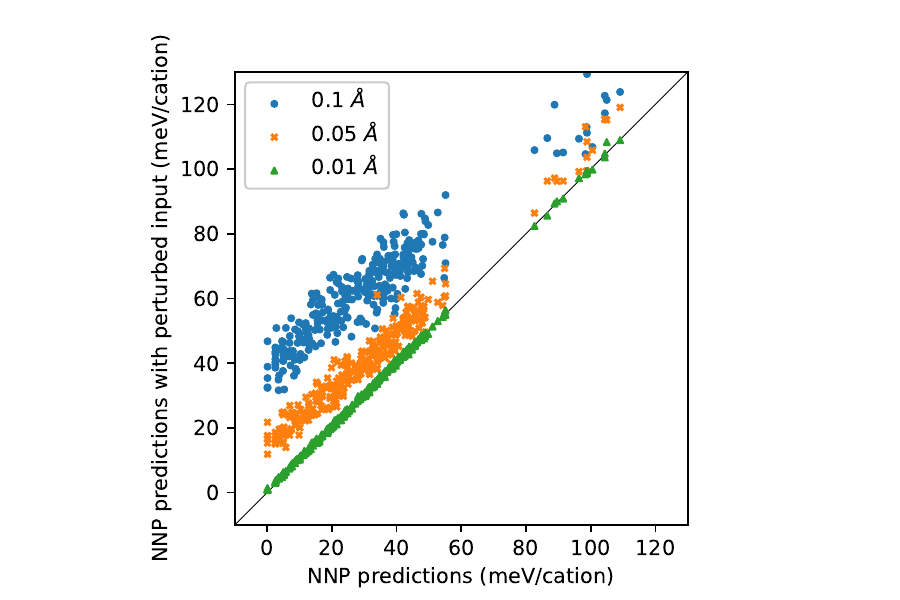}
    \caption{Predicted NNP energies when the input coordinates of all atoms are
    perturbed in random directions with a magnitude of 0.1 \r{A} (blue circles),
    0.05 \r{A} (orange crosses), and 0.01 \r{A} (green triangles). The results are
    plotted against the predictions using the unperturbed ideal lattice coordinates as input.
    }
    \label{fig:NNPperturb}
\end{figure}

As noted in Sec.~\ref{Sec:NNPdetails}, we designed our NNP model to be trained
and used only on coordinates of the ideal lattice $\vec{\sigma} \in
\{\vec{\sigma}_\text{lattice}\}$. However, there is no way to specify the {\it
exact} lattice coordinate due to the nature of floating point numbers used to
represent non-integer values in computers. Thus, to see how robust our method
is against deviations in the input coordinates, 
we examined the predictions of our final NNP model for 
randomly perturbed lattices (Fig.~\ref{fig:NNPperturb}). We find that the
predictions are virtually indistinguishable from predictions on the corresponding
perfect lattice structures (that
is, perfect within the numerical precision of our software stack) when 
each atom is perturbed by 0.01 \r{A}. The NNP looks usable up to a perturbation of
$\sim 0.05$ \r{A}, while at 0.1 \r{A}, we start to observe sizable deviations.
0.01 \r{A} is not that small, and it should not be too difficult to make sure
that the errors in the input coordinates fall within this range.
The robustness may originate from the fact that 
the NNP maps similar input to similar energies, and because there are no other
similar structures to choose from in the training set space except the
corresponding ideal lattice structure.

\subsection{Applicability to hard-to-relax or frustrated systems}
Finally, we note that for truly complicated systems, local frustrations may lead to
several local minima when starting the relaxation from ideal lattice structures. This
means that the implicitly assumed one-to-one correspondence between lattice
configuration and relaxed energy is broken. In such
cases, it may be advisable to perform the active learning scheme for a few
iterations, then use the final NNP in nested Monte Carlo
\cite{iftimie2000,Hetenyi2002,gelb2003,Leiding2016,nagai2020,Nagai2020b} or upsampling 
schemes \cite{zhang2018}, where MC or MD steps are carried out with 
low cost potentials then augmented with DFT calculations at preset intervals.
These schemes are guaranteed to converge to the same result as sampling based only on DFT
energies, and the speedup will depend on the quality of the low cost potential.
The frustration problem mentioned above can be solved by nested DFT
relaxations during the MC sampling: by randomly perturbing the structure
slightly before relaxation, the system has a chance of visiting any of those
local minima. This is in the spirit of basin-hopping MC \cite{Wales1997}.
We plan on implementing and testing such nesting schemes in the near future.

% , although
% the current work estimates the DOI above 600 K to be slightly larger than
% previous works. 
% It is difficult to judge whether the current work is more
% accurate than the CE-SPCM approach of Ref.~\citenum{Seko2014}. The learning
% is performed in a much larger supercell here (192 cation model vs.~18--30 cation
% model in Ref.~\citenum{Seko2014}), and the resulting prediction error of a few
% meV/cation (Fig.~\ref{fig:ALcheck}) is smaller than $\sim$10 meV/cation reported in
% Ref.~\citenum{Seko2014}. In that sense, our approach seems to be a clear improvement over
% cluster expansion, and it is also much easier to use as discussed above. On the
% other hand, the MC calculations are performed in a 
% much smaller cell (192 vs. 48 000). Unfortunately, NNP evaluation is not as fast as
% cluster expansion at this point, and it is not fast enough to enable sufficient sampling on a
% 48 000-cation model. Thus, we cannot conclude at this point whether the slight
% discrepancy in Fig.~\ref{fig:DOI} compared to CE-SPCM is due to higher accuracy
% or poorer sampling. 

\section{Summary}
In this work, we proposed a new method to obtain lightweight models that can be
used to sample millions of configurations in the {\it ab initio} lattice
configuration problem:
Behler-Parinello NNPs are trained to predict relaxed energies from perfect-lattice
structures with configurational disorder, and the NNPs are iteratively refined
using an active learning scheme involving several RXMC runs. The idea was tested
successfully in calculating the degree of A/B site inversion in three spinel oxides: 
MgAl$_2$O$_4$, ZnAl$_2$O$_4$, and MgGa$_2$O$_4$.
Merits compared to conventional continuous-coordinate NNP models include the
ability to bypass structural relaxation, as well as the restriction on the
input coordinate space that makes machine learning much easier.
This approach may also be considered a `lazy' alternative to cluster expansion; no
explicit cluster selection is necessary, and it also avoids the combinatorial 
explosion in the amount of computation which makes cluster expansion practically 
unusable for 
many-component systems with multiple sublattices. 
We also demonstrated that the scheme is at least as robust as cluster expansion
with respect to magnitude of local relaxation, and also in treating systems with
long-range interactions.
Due to the merits outlined above, we believe that this approach
will enable configurational sampling in complex bulk or interface systems where
it is technically difficult, if not impossible, to apply other previous approaches. 

% In addition, the importance of having large enough supercells in 
% the training set was highlighted. This applies equally to any model based on
% local decomposition of the total energy including cluster expansion.
% The fact that convergence was seen within 1000 atoms in 
% a multivalent ionic system where one would expect significant long-range
% interactions gives hope; since 1000-atom calculations are within reach
% of plane-wave DFT calculations using modern day supercomputer resources,
% we advocate testing for convergence with explicit
% DFT calculations on larger supercells whenever possible.

\acknowledgments
The software used in
this work (abICS) was developed in part by ``Project for advancement of software usability
in materials science'' by the Institute for Solid State Physics, the University of
Tokyo, and the calculations were performed on the joint-use supercomputer system
at the same Institute. This research was supported by ``Program for Promoting
Research on the Supercomputer Fugaku'' (Fugaku Battery \&
Fuel Cell Project), and by JST CREST Program (Grant Number JPMJCR18J3, Japan). 
S.~K.~acknowledges funding from Japan Society for
the Promotion of Science (JSPS) KAKENHI Grant Number 20H05284 (Grant-in-Aid for
Scientific Research on Innovative Areas ``Interface IONICS''), 
as well as support from JST FOREST Program
(Grant Number JPMJFR2037, Japan). T.~O. is funded by JSPS KAKENHI Grant
Number 19H05792 (Grant-in-Aid for Scientific Research on Innovative Areas
``Crystal Defect Core'') and JSPS KAKENHI Grant Number 21K04648 (Grain-in-Aid for Scientific Research(C)).
Plots of the data were generated using Matplotlib \cite{Hunter2007}
and seaborn \cite{Waskom2021}.

% \section{Author Contributions}
% S.~K. and T.~O. designed the research. U.~M. performed
% preliminary NNP fitting calculations to assess the feasibility of the approach.
% S.~K., K.~Y., and Y.~M. jointly coded the abICS software. S.~K. performed all of
% the calculations and analyses presented in this paper. All authors contributed
% to the discussion and steering of the research direction and jointly wrote the paper.

% \section{Additional information}
% \subsection{Competing interests}
% The Authors declare no Competing Financial or Non-Financial Interests.
% active learning: step 1: 300 structures from 6400 RXMC steps (15 replicas x 20
% samples)
% step 2 and later: 300 structures from 16000 RXMC steps (15 replicas x 20
% samples)
% DOI calculation: 
% step 0: 500-1000 (x16)
% step 1: q

\bibliography{article}

\begin{thebibliography}{72}%
\makeatletter
\providecommand \@ifxundefined [1]{%
 \@ifx{#1\undefined}
}%
\providecommand \@ifnum [1]{%
 \ifnum #1\expandafter \@firstoftwo
 \else \expandafter \@secondoftwo
 \fi
}%
\providecommand \@ifx [1]{%
 \ifx #1\expandafter \@firstoftwo
 \else \expandafter \@secondoftwo
 \fi
}%
\providecommand \natexlab [1]{#1}%
\providecommand \enquote  [1]{``#1''}%
\providecommand \bibnamefont  [1]{#1}%
\providecommand \bibfnamefont [1]{#1}%
\providecommand \citenamefont [1]{#1}%
\providecommand \href@noop [0]{\@secondoftwo}%
\providecommand \href [0]{\begingroup \@sanitize@url \@href}%
\providecommand \@href[1]{\@@startlink{#1}\@@href}%
\providecommand \@@href[1]{\endgroup#1\@@endlink}%
\providecommand \@sanitize@url [0]{\catcode `\\12\catcode `\$12\catcode
  `\&12\catcode `\#12\catcode `\^12\catcode `\_12\catcode `\%12\relax}%
\providecommand \@@startlink[1]{}%
\providecommand \@@endlink[0]{}%
\providecommand \url  [0]{\begingroup\@sanitize@url \@url }%
\providecommand \@url [1]{\endgroup\@href {#1}{\urlprefix }}%
\providecommand \urlprefix  [0]{URL }%
\providecommand \Eprint [0]{\href }%
\providecommand \doibase [0]{http://dx.doi.org/}%
\providecommand \selectlanguage [0]{\@gobble}%
\providecommand \bibinfo  [0]{\@secondoftwo}%
\providecommand \bibfield  [0]{\@secondoftwo}%
\providecommand \translation [1]{[#1]}%
\providecommand \BibitemOpen [0]{}%
\providecommand \bibitemStop [0]{}%
\providecommand \bibitemNoStop [0]{.\EOS\space}%
\providecommand \EOS [0]{\spacefactor3000\relax}%
\providecommand \BibitemShut  [1]{\csname bibitem#1\endcsname}%
\let\auto@bib@innerbib\@empty
%</preamble>
\bibitem [{\citenamefont {Hafner}, \citenamefont {Wolverton},\ and\
  \citenamefont {Ceder}(2006)}]{Hafner2006}%
  \BibitemOpen
  \bibfield  {author} {\bibinfo {author} {\bibfnamefont {J.}~\bibnamefont
  {Hafner}}, \bibinfo {author} {\bibfnamefont {C.}~\bibnamefont {Wolverton}}, \
  and\ \bibinfo {author} {\bibfnamefont {G.}~\bibnamefont {Ceder}},\ }\bibfield
   {title} {\enquote {\bibinfo {title} {{Toward Computational Materials Design
  : The Impact of Density Functional Theory of Materials Research}},}\
  }\href@noop {} {\bibfield  {journal} {\bibinfo  {journal} {MRS Bull.}\
  }\textbf {\bibinfo {volume} {31}},\ \bibinfo {pages} {659--668} (\bibinfo
  {year} {2006})}\BibitemShut {NoStop}%
\bibitem [{\citenamefont {Ceder}(2010)}]{Ceder2010}%
  \BibitemOpen
  \bibfield  {author} {\bibinfo {author} {\bibfnamefont {G.}~\bibnamefont
  {Ceder}},\ }\bibfield  {title} {\enquote {\bibinfo {title} {{Opportunities
  and challenges for first-principles materials design and applications to Li
  battery materials}},}\ }\href@noop {} {\bibfield  {journal} {\bibinfo
  {journal} {MRS Bull.}\ }\textbf {\bibinfo {volume} {35}},\ \bibinfo {pages}
  {693--701} (\bibinfo {year} {2010})}\BibitemShut {NoStop}%
\bibitem [{\citenamefont {Neugebauer}\ and\ \citenamefont
  {Hickel}(2013)}]{neugebauer2013}%
  \BibitemOpen
  \bibfield  {author} {\bibinfo {author} {\bibfnamefont {J.}~\bibnamefont
  {Neugebauer}}\ and\ \bibinfo {author} {\bibfnamefont {T.}~\bibnamefont
  {Hickel}},\ }\bibfield  {title} {\enquote {\bibinfo {title} {Density
  functional theory in materials science},}\ }\href {\doibase
  10.1002/wcms.1125} {\bibfield  {journal} {\bibinfo  {journal} {WIREs Comput.
  Mol. Sci.}\ }\textbf {\bibinfo {volume} {3}},\ \bibinfo {pages} {438--448}
  (\bibinfo {year} {2013})}\BibitemShut {NoStop}%
\bibitem [{\citenamefont {Sanchez}, \citenamefont {Ducastelle},\ and\
  \citenamefont {Gratias}(1984)}]{Sanchez1984}%
  \BibitemOpen
  \bibfield  {author} {\bibinfo {author} {\bibfnamefont {J.~M.}\ \bibnamefont
  {Sanchez}}, \bibinfo {author} {\bibfnamefont {F.}~\bibnamefont {Ducastelle}},
  \ and\ \bibinfo {author} {\bibfnamefont {D.}~\bibnamefont {Gratias}},\
  }\bibfield  {title} {\enquote {\bibinfo {title} {{Generalized cluster
  description of multicomponent systems}},}\ }\href@noop {} {\bibfield
  {journal} {\bibinfo  {journal} {Physica A}\ }\textbf {\bibinfo {volume}
  {128A}},\ \bibinfo {pages} {334--350} (\bibinfo {year} {1984})}\BibitemShut
  {NoStop}%
\bibitem [{\citenamefont {Ceder}(1993)}]{Ceder1993}%
  \BibitemOpen
  \bibfield  {author} {\bibinfo {author} {\bibfnamefont {G.}~\bibnamefont
  {Ceder}},\ }\bibfield  {title} {\enquote {\bibinfo {title} {A derivation of
  the {Ising} model for the computation of phase diagrams},}\ }\href {\doibase
  10.1016/0927-0256(93)90005-8} {\bibfield  {journal} {\bibinfo  {journal}
  {Comput. Mater. Sci.}\ }\textbf {\bibinfo {volume} {1}},\ \bibinfo {pages}
  {144--150} (\bibinfo {year} {1993})}\BibitemShut {NoStop}%
\bibitem [{\citenamefont {van~de Walle}, \citenamefont {Asta},\ and\
  \citenamefont {Ceder}(2002)}]{Walle2002}%
  \BibitemOpen
  \bibfield  {author} {\bibinfo {author} {\bibfnamefont {A.}~\bibnamefont
  {van~de Walle}}, \bibinfo {author} {\bibfnamefont {M.}~\bibnamefont {Asta}},
  \ and\ \bibinfo {author} {\bibfnamefont {G.}~\bibnamefont {Ceder}},\
  }\bibfield  {title} {\enquote {\bibinfo {title} {The alloy theoretic
  automated toolkit: A user guide},}\ }\href {\doibase
  https://doi.org/10.1016/S0364-5916(02)80006-2} {\bibfield  {journal}
  {\bibinfo  {journal} {Calphad}\ }\textbf {\bibinfo {volume} {26}},\ \bibinfo
  {pages} {539 -- 553} (\bibinfo {year} {2002})}\BibitemShut {NoStop}%
\bibitem [{\citenamefont {Seko}, \citenamefont {Koyama},\ and\ \citenamefont
  {Tanaka}(2009)}]{Seko2009}%
  \BibitemOpen
  \bibfield  {author} {\bibinfo {author} {\bibfnamefont {A.}~\bibnamefont
  {Seko}}, \bibinfo {author} {\bibfnamefont {Y.}~\bibnamefont {Koyama}}, \ and\
  \bibinfo {author} {\bibfnamefont {I.}~\bibnamefont {Tanaka}},\ }\bibfield
  {title} {\enquote {\bibinfo {title} {Cluster expansion method for
  multicomponent systems based on optimal selection of structures for
  density-functional theory calculations},}\ }\href {\doibase
  10.1103/PhysRevB.80.165122} {\bibfield  {journal} {\bibinfo  {journal} {Phys.
  Rev. B}\ }\textbf {\bibinfo {volume} {80}},\ \bibinfo {pages} {165122}
  (\bibinfo {year} {2009})}\BibitemShut {NoStop}%
\bibitem [{\citenamefont {Sanchez}(2010)}]{Sanchez2010}%
  \BibitemOpen
  \bibfield  {author} {\bibinfo {author} {\bibfnamefont {J.~M.}\ \bibnamefont
  {Sanchez}},\ }\bibfield  {title} {\enquote {\bibinfo {title} {Cluster
  expansion and the configurational theory of alloys},}\ }\href {\doibase
  10.1103/PhysRevB.81.224202} {\bibfield  {journal} {\bibinfo  {journal} {Phys.
  Rev. B}\ }\textbf {\bibinfo {volume} {81}},\ \bibinfo {pages} {224202}
  (\bibinfo {year} {2010})}\BibitemShut {NoStop}%
\bibitem [{\citenamefont {Sanchez}(2017)}]{Sanchez2017}%
  \BibitemOpen
  \bibfield  {author} {\bibinfo {author} {\bibfnamefont {J.~M.}\ \bibnamefont
  {Sanchez}},\ }\bibfield  {title} {\enquote {\bibinfo {title} {{Foundations
  and Practical Implementations of the Cluster Expansion}},}\ }\href@noop {}
  {\bibfield  {journal} {\bibinfo  {journal} {J. Phase Equilib. Diffus.}\
  }\textbf {\bibinfo {volume} {38}},\ \bibinfo {pages} {238--251} (\bibinfo
  {year} {2017})}\BibitemShut {NoStop}%
\bibitem [{\citenamefont {Wu}\ \emph {et~al.}(2016)\citenamefont {Wu},
  \citenamefont {He}, \citenamefont {Song}, \citenamefont {Gao},\ and\
  \citenamefont {Shi}}]{Wu2016}%
  \BibitemOpen
  \bibfield  {author} {\bibinfo {author} {\bibfnamefont {Q.}~\bibnamefont
  {Wu}}, \bibinfo {author} {\bibfnamefont {B.}~\bibnamefont {He}}, \bibinfo
  {author} {\bibfnamefont {T.}~\bibnamefont {Song}}, \bibinfo {author}
  {\bibfnamefont {J.}~\bibnamefont {Gao}}, \ and\ \bibinfo {author}
  {\bibfnamefont {S.}~\bibnamefont {Shi}},\ }\bibfield  {title} {\enquote
  {\bibinfo {title} {{Cluster expansion method and its application in
  computational materials science}},}\ }\href@noop {} {\bibfield  {journal}
  {\bibinfo  {journal} {Comput. Mater. Sci.}\ }\textbf {\bibinfo {volume}
  {125}},\ \bibinfo {pages} {243--254} (\bibinfo {year} {2016})}\BibitemShut
  {NoStop}%
\bibitem [{\citenamefont {Chang}\ \emph {et~al.}(2019)\citenamefont {Chang},
  \citenamefont {Kleiven}, \citenamefont {Melander}, \citenamefont {Akola},
  \citenamefont {Garcia-Lastra},\ and\ \citenamefont {Vegge}}]{Chang2019}%
  \BibitemOpen
  \bibfield  {author} {\bibinfo {author} {\bibfnamefont {J.~H.}\ \bibnamefont
  {Chang}}, \bibinfo {author} {\bibfnamefont {D.}~\bibnamefont {Kleiven}},
  \bibinfo {author} {\bibfnamefont {M.}~\bibnamefont {Melander}}, \bibinfo
  {author} {\bibfnamefont {J.}~\bibnamefont {Akola}}, \bibinfo {author}
  {\bibfnamefont {J.~M.}\ \bibnamefont {Garcia-Lastra}}, \ and\ \bibinfo
  {author} {\bibfnamefont {T.}~\bibnamefont {Vegge}},\ }\bibfield  {title}
  {\enquote {\bibinfo {title} {{CLEASE}: a versatile and user-friendly
  implementation of cluster expansion method},}\ }\href {\doibase
  10.1088/1361-648X/ab1bbc} {\bibfield  {journal} {\bibinfo  {journal} {J.
  Phys.: Condens. Matter}\ }\textbf {\bibinfo {volume} {31}},\ \bibinfo {pages}
  {325901} (\bibinfo {year} {2019})}\BibitemShut {NoStop}%
\bibitem [{\citenamefont {Nguyen}\ \emph {et~al.}(2017)\citenamefont {Nguyen},
  \citenamefont {Rosenbrock}, \citenamefont {Reese},\ and\ \citenamefont
  {Hart}}]{Nguyen2017}%
  \BibitemOpen
  \bibfield  {author} {\bibinfo {author} {\bibfnamefont {A.~H.}\ \bibnamefont
  {Nguyen}}, \bibinfo {author} {\bibfnamefont {C.~W.}\ \bibnamefont
  {Rosenbrock}}, \bibinfo {author} {\bibfnamefont {C.~S.}\ \bibnamefont
  {Reese}}, \ and\ \bibinfo {author} {\bibfnamefont {G.~L.~W.}\ \bibnamefont
  {Hart}},\ }\bibfield  {title} {\enquote {\bibinfo {title} {{Robustness of the
  cluster expansion: Assessing the roles of relaxation and numerical error}},}\
  }\href@noop {} {\bibfield  {journal} {\bibinfo  {journal} {Phys. Rev. B}\
  }\textbf {\bibinfo {volume} {96}},\ \bibinfo {pages} {014107} (\bibinfo
  {year} {2017})}\BibitemShut {NoStop}%
\bibitem [{\citenamefont {Seko}\ and\ \citenamefont {Tanaka}(2014)}]{Seko2014}%
  \BibitemOpen
  \bibfield  {author} {\bibinfo {author} {\bibfnamefont {A.}~\bibnamefont
  {Seko}}\ and\ \bibinfo {author} {\bibfnamefont {I.}~\bibnamefont {Tanaka}},\
  }\bibfield  {title} {\enquote {\bibinfo {title} {{Cluster expansion of
  multicomponent ionic systems with controlled accuracy: importance of
  long-range interactions in heterovalent ionic systems}},}\ }\href@noop {}
  {\bibfield  {journal} {\bibinfo  {journal} {J. Phys.: Condens. Matter}\
  }\textbf {\bibinfo {volume} {26}},\ \bibinfo {pages} {115403} (\bibinfo
  {year} {2014})}\BibitemShut {NoStop}%
\bibitem [{\citenamefont {Seko}\ and\ \citenamefont {Tanaka}(2011)}]{Seko2011}%
  \BibitemOpen
  \bibfield  {author} {\bibinfo {author} {\bibfnamefont {A.}~\bibnamefont
  {Seko}}\ and\ \bibinfo {author} {\bibfnamefont {I.}~\bibnamefont {Tanaka}},\
  }\bibfield  {title} {\enquote {\bibinfo {title} {Grouping of structures for
  cluster expansion of multicomponent systems with controlled accuracy},}\
  }\href {\doibase 10.1103/PhysRevB.83.224111} {\bibfield  {journal} {\bibinfo
  {journal} {Phys. Rev. B}\ }\textbf {\bibinfo {volume} {83}},\ \bibinfo
  {pages} {224111} (\bibinfo {year} {2011})}\BibitemShut {NoStop}%
\bibitem [{\citenamefont {Leong}\ and\ \citenamefont {Tan}(2019)}]{Leong2019}%
  \BibitemOpen
  \bibfield  {author} {\bibinfo {author} {\bibfnamefont {Z.}~\bibnamefont
  {Leong}}\ and\ \bibinfo {author} {\bibfnamefont {T.~L.}\ \bibnamefont
  {Tan}},\ }\bibfield  {title} {\enquote {\bibinfo {title} {Robust cluster
  expansion of multicomponent systems using structured sparsity},}\ }\href
  {\doibase 10.1103/PhysRevB.100.134108} {\bibfield  {journal} {\bibinfo
  {journal} {Phys. Rev. B}\ }\textbf {\bibinfo {volume} {100}},\ \bibinfo
  {pages} {134108} (\bibinfo {year} {2019})}\BibitemShut {NoStop}%
\bibitem [{\citenamefont {Khan}\ and\ \citenamefont
  {Eisenbach}(2016)}]{Khan2016}%
  \BibitemOpen
  \bibfield  {author} {\bibinfo {author} {\bibfnamefont {S.~N.}\ \bibnamefont
  {Khan}}\ and\ \bibinfo {author} {\bibfnamefont {M.}~\bibnamefont
  {Eisenbach}},\ }\bibfield  {title} {\enquote {\bibinfo {title}
  {{Density-functional Monte-Carlo simulation of CuZn order-disorder
  transition}},}\ }\href {\doibase 10.1103/PhysRevB.93.024203} {\bibfield
  {journal} {\bibinfo  {journal} {Phys. Rev. B}\ }\textbf {\bibinfo {volume}
  {93}},\ \bibinfo {pages} {024203} (\bibinfo {year} {2016})}\BibitemShut
  {NoStop}%
\bibitem [{\citenamefont {Wexler}, \citenamefont {Qiu},\ and\ \citenamefont
  {Rappe}(2019)}]{Wexler2019}%
  \BibitemOpen
  \bibfield  {author} {\bibinfo {author} {\bibfnamefont {R.~B.}\ \bibnamefont
  {Wexler}}, \bibinfo {author} {\bibfnamefont {T.}~\bibnamefont {Qiu}}, \ and\
  \bibinfo {author} {\bibfnamefont {A.~M.}\ \bibnamefont {Rappe}},\ }\bibfield
  {title} {\enquote {\bibinfo {title} {{Automatic Prediction of Surface Phase
  Diagrams Using Ab Initio Grand Canonical Monte Carlo}},}\ }\href {\doibase
  10.1021/acs.jpcc.8b11093} {\bibfield  {journal} {\bibinfo  {journal} {J.
  Phys. Chem. C}\ }\textbf {\bibinfo {volume} {123}},\ \bibinfo {pages}
  {2321--2328} (\bibinfo {year} {2019})}\BibitemShut {NoStop}%
\bibitem [{\citenamefont {Kasamatsu}\ and\ \citenamefont
  {Sugino}(2019)}]{Kasamatsu2019}%
  \BibitemOpen
  \bibfield  {author} {\bibinfo {author} {\bibfnamefont {S.}~\bibnamefont
  {Kasamatsu}}\ and\ \bibinfo {author} {\bibfnamefont {O.}~\bibnamefont
  {Sugino}},\ }\bibfield  {title} {\enquote {\bibinfo {title} {{Direct coupling
  of first-principles calculations with replica exchange Monte Carlo sampling
  of ion disorder in solids}},}\ }\href {\doibase 10.1088/1361-648X/aaf75c}
  {\bibfield  {journal} {\bibinfo  {journal} {J. Phys.: Condens. Matter}\
  }\textbf {\bibinfo {volume} {31}},\ \bibinfo {pages} {085901} (\bibinfo
  {year} {2019})}\BibitemShut {NoStop}%
\bibitem [{\citenamefont {Kasamatsu}\ \emph {et~al.}(2020)\citenamefont
  {Kasamatsu}, \citenamefont {Sugino}, \citenamefont {Ogawa},\ and\
  \citenamefont {Kuwabara}}]{Kasamatsu2020}%
  \BibitemOpen
  \bibfield  {author} {\bibinfo {author} {\bibfnamefont {S.}~\bibnamefont
  {Kasamatsu}}, \bibinfo {author} {\bibfnamefont {O.}~\bibnamefont {Sugino}},
  \bibinfo {author} {\bibfnamefont {T.}~\bibnamefont {Ogawa}}, \ and\ \bibinfo
  {author} {\bibfnamefont {A.}~\bibnamefont {Kuwabara}},\ }\bibfield  {title}
  {\enquote {\bibinfo {title} {Dopant arrangements in {Y}-doped {BaZrO}$_3$
  under processing conditions and their impact on proton conduction: a
  large-scale first-principles thermodynamics study},}\ }\href {\doibase
  10.1039/D0TA01741H} {\bibfield  {journal} {\bibinfo  {journal} {J. Mater.
  Chem. A}\ }\textbf {\bibinfo {volume} {8}},\ \bibinfo {pages} {12674--12686}
  (\bibinfo {year} {2020})}\BibitemShut {NoStop}%
\bibitem [{\citenamefont {Wang}\ and\ \citenamefont {Landau}(2001)}]{Wang2001}%
  \BibitemOpen
  \bibfield  {author} {\bibinfo {author} {\bibfnamefont {F.}~\bibnamefont
  {Wang}}\ and\ \bibinfo {author} {\bibfnamefont {D.~P.}\ \bibnamefont
  {Landau}},\ }\bibfield  {title} {\enquote {\bibinfo {title} {{Efficient,
  multiple-range random walk algorithm to calculate the density of states}},}\
  }\href@noop {} {\bibfield  {journal} {\bibinfo  {journal} {Phys. Rev. Lett.}\
  }\textbf {\bibinfo {volume} {86}},\ \bibinfo {pages} {2050--2053} (\bibinfo
  {year} {2001})}\BibitemShut {NoStop}%
\bibitem [{\citenamefont {Hukushima}\ and\ \citenamefont
  {Nemoto}(1996)}]{Hukushima1996}%
  \BibitemOpen
  \bibfield  {author} {\bibinfo {author} {\bibfnamefont {K.}~\bibnamefont
  {Hukushima}}\ and\ \bibinfo {author} {\bibfnamefont {K.}~\bibnamefont
  {Nemoto}},\ }\bibfield  {title} {\enquote {\bibinfo {title} {{Exchange Monte
  Carlo Method and Application to Spin Glass Simulations}},}\ }\href@noop {}
  {\bibfield  {journal} {\bibinfo  {journal} {J. Phys. Soc. Jpn.}\ }\textbf
  {\bibinfo {volume} {65}},\ \bibinfo {pages} {1604--1608} (\bibinfo {year}
  {1996})}\BibitemShut {NoStop}%
\bibitem [{\citenamefont {Podryabinkin}\ and\ \citenamefont
  {Shapeev}(2017)}]{Podryabinkin2017}%
  \BibitemOpen
  \bibfield  {author} {\bibinfo {author} {\bibfnamefont {E.~V.}\ \bibnamefont
  {Podryabinkin}}\ and\ \bibinfo {author} {\bibfnamefont {A.~V.}\ \bibnamefont
  {Shapeev}},\ }\bibfield  {title} {\enquote {\bibinfo {title} {Active learning
  of linearly parametrized interatomic potentials},}\ }\href {\doibase
  10.1016/j.commatsci.2017.08.031} {\bibfield  {journal} {\bibinfo  {journal}
  {Comput. Mater. Sci.}\ }\textbf {\bibinfo {volume} {140}},\ \bibinfo {pages}
  {171--180} (\bibinfo {year} {2017})}\BibitemShut {NoStop}%
\bibitem [{\citenamefont {Podryabinkin}\ \emph {et~al.}(2019)\citenamefont
  {Podryabinkin}, \citenamefont {Tikhonov}, \citenamefont {Shapeev},\ and\
  \citenamefont {Oganov}}]{Podryabinkin2019}%
  \BibitemOpen
  \bibfield  {author} {\bibinfo {author} {\bibfnamefont {E.~V.}\ \bibnamefont
  {Podryabinkin}}, \bibinfo {author} {\bibfnamefont {E.~V.}\ \bibnamefont
  {Tikhonov}}, \bibinfo {author} {\bibfnamefont {A.~V.}\ \bibnamefont
  {Shapeev}}, \ and\ \bibinfo {author} {\bibfnamefont {A.~R.}\ \bibnamefont
  {Oganov}},\ }\bibfield  {title} {\enquote {\bibinfo {title} {Accelerating
  crystal structure prediction by machine-learning interatomic potentials with
  active learning},}\ }\href {\doibase 10.1103/PhysRevB.99.064114} {\bibfield
  {journal} {\bibinfo  {journal} {Phys. Rev. B}\ }\textbf {\bibinfo {volume}
  {99}},\ \bibinfo {pages} {064114} (\bibinfo {year} {2019})}\BibitemShut
  {NoStop}%
\bibitem [{\citenamefont {Gubaev}\ \emph {et~al.}(2019)\citenamefont {Gubaev},
  \citenamefont {Podryabinkin}, \citenamefont {Hart},\ and\ \citenamefont
  {Shapeev}}]{Gubaev2019}%
  \BibitemOpen
  \bibfield  {author} {\bibinfo {author} {\bibfnamefont {K.}~\bibnamefont
  {Gubaev}}, \bibinfo {author} {\bibfnamefont {E.~V.}\ \bibnamefont
  {Podryabinkin}}, \bibinfo {author} {\bibfnamefont {G.~L.}\ \bibnamefont
  {Hart}}, \ and\ \bibinfo {author} {\bibfnamefont {A.~V.}\ \bibnamefont
  {Shapeev}},\ }\bibfield  {title} {\enquote {\bibinfo {title} {Accelerating
  high-throughput searches for new alloys with active learning of interatomic
  potentials},}\ }\href {\doibase 10.1016/j.commatsci.2018.09.031} {\bibfield
  {journal} {\bibinfo  {journal} {Comput. Mater. Sci.}\ }\textbf {\bibinfo
  {volume} {156}},\ \bibinfo {pages} {148--156} (\bibinfo {year}
  {2019})}\BibitemShut {NoStop}%
\bibitem [{\citenamefont {Smith}\ \emph {et~al.}(2018)\citenamefont {Smith},
  \citenamefont {Nebgen}, \citenamefont {Lubbers}, \citenamefont {Isayev},\
  and\ \citenamefont {Roitberg}}]{Smith2018}%
  \BibitemOpen
  \bibfield  {author} {\bibinfo {author} {\bibfnamefont {J.~S.}\ \bibnamefont
  {Smith}}, \bibinfo {author} {\bibfnamefont {B.}~\bibnamefont {Nebgen}},
  \bibinfo {author} {\bibfnamefont {N.}~\bibnamefont {Lubbers}}, \bibinfo
  {author} {\bibfnamefont {O.}~\bibnamefont {Isayev}}, \ and\ \bibinfo {author}
  {\bibfnamefont {A.~E.}\ \bibnamefont {Roitberg}},\ }\bibfield  {title}
  {\enquote {\bibinfo {title} {Less is more: {Sampling} chemical space with
  active learning},}\ }\href {\doibase 10.1063/1.5023802} {\bibfield  {journal}
  {\bibinfo  {journal} {J. Chem. Phys.}\ }\textbf {\bibinfo {volume} {148}},\
  \bibinfo {pages} {241733} (\bibinfo {year} {2018})}\BibitemShut {NoStop}%
\bibitem [{\citenamefont {Zhang}\ \emph {et~al.}(2019)\citenamefont {Zhang},
  \citenamefont {Lin}, \citenamefont {Wang}, \citenamefont {Car},\ and\
  \citenamefont {E}}]{Zhang2019}%
  \BibitemOpen
  \bibfield  {author} {\bibinfo {author} {\bibfnamefont {L.}~\bibnamefont
  {Zhang}}, \bibinfo {author} {\bibfnamefont {D.-Y.}\ \bibnamefont {Lin}},
  \bibinfo {author} {\bibfnamefont {H.}~\bibnamefont {Wang}}, \bibinfo {author}
  {\bibfnamefont {R.}~\bibnamefont {Car}}, \ and\ \bibinfo {author}
  {\bibfnamefont {W.}~\bibnamefont {E}},\ }\bibfield  {title} {\enquote
  {\bibinfo {title} {Active learning of uniformly accurate interatomic
  potentials for materials simulation},}\ }\href {\doibase
  10.1103/PhysRevMaterials.3.023804} {\bibfield  {journal} {\bibinfo  {journal}
  {Phys. Rev. Mater.}\ }\textbf {\bibinfo {volume} {3}},\ \bibinfo {pages}
  {023804} (\bibinfo {year} {2019})}\BibitemShut {NoStop}%
\bibitem [{\citenamefont {Kostiuchenko}\ \emph {et~al.}(2019)\citenamefont
  {Kostiuchenko}, \citenamefont {Körmann}, \citenamefont {Neugebauer},\ and\
  \citenamefont {Shapeev}}]{Kostiuchenko2019}%
  \BibitemOpen
  \bibfield  {author} {\bibinfo {author} {\bibfnamefont {T.}~\bibnamefont
  {Kostiuchenko}}, \bibinfo {author} {\bibfnamefont {F.}~\bibnamefont
  {Körmann}}, \bibinfo {author} {\bibfnamefont {J.}~\bibnamefont
  {Neugebauer}}, \ and\ \bibinfo {author} {\bibfnamefont {A.}~\bibnamefont
  {Shapeev}},\ }\bibfield  {title} {\enquote {\bibinfo {title} {Impact of
  lattice relaxations on phase transitions in a high-entropy alloy studied by
  machine-learning potentials},}\ }\href {\doibase 10.1038/s41524-019-0195-y}
  {\bibfield  {journal} {\bibinfo  {journal} {npj Comput. Mater.}\ }\textbf
  {\bibinfo {volume} {5}},\ \bibinfo {pages} {1--7} (\bibinfo {year}
  {2019})}\BibitemShut {NoStop}%
\bibitem [{\citenamefont {Loeffler}\ \emph {et~al.}(2020)\citenamefont
  {Loeffler}, \citenamefont {Patra}, \citenamefont {Chan}, \citenamefont
  {Cherukara},\ and\ \citenamefont {Sankaranarayanan}}]{Loeffler2020}%
  \BibitemOpen
  \bibfield  {author} {\bibinfo {author} {\bibfnamefont {T.~D.}\ \bibnamefont
  {Loeffler}}, \bibinfo {author} {\bibfnamefont {T.~K.}\ \bibnamefont {Patra}},
  \bibinfo {author} {\bibfnamefont {H.}~\bibnamefont {Chan}}, \bibinfo {author}
  {\bibfnamefont {M.}~\bibnamefont {Cherukara}}, \ and\ \bibinfo {author}
  {\bibfnamefont {S.~K. R.~S.}\ \bibnamefont {Sankaranarayanan}},\ }\bibfield
  {title} {\enquote {\bibinfo {title} {Active {Learning} the {Potential}
  {Energy} {Landscape} for {Water} {Clusters} from {Sparse} {Training}
  {Data}},}\ }\href {\doibase 10.1021/acs.jpcc.0c00047} {\bibfield  {journal}
  {\bibinfo  {journal} {J. Phys. Chem. C}\ }\textbf {\bibinfo {volume} {124}},\
  \bibinfo {pages} {4907--4916} (\bibinfo {year} {2020})}\BibitemShut {NoStop}%
\bibitem [{\citenamefont {Vandermause}\ \emph {et~al.}(2020)\citenamefont
  {Vandermause}, \citenamefont {Torrisi}, \citenamefont {Batzner},
  \citenamefont {Xie}, \citenamefont {Sun}, \citenamefont {Kolpak},\ and\
  \citenamefont {Kozinsky}}]{Vandermause2020}%
  \BibitemOpen
  \bibfield  {author} {\bibinfo {author} {\bibfnamefont {J.}~\bibnamefont
  {Vandermause}}, \bibinfo {author} {\bibfnamefont {S.~B.}\ \bibnamefont
  {Torrisi}}, \bibinfo {author} {\bibfnamefont {S.}~\bibnamefont {Batzner}},
  \bibinfo {author} {\bibfnamefont {Y.}~\bibnamefont {Xie}}, \bibinfo {author}
  {\bibfnamefont {L.}~\bibnamefont {Sun}}, \bibinfo {author} {\bibfnamefont
  {A.~M.}\ \bibnamefont {Kolpak}}, \ and\ \bibinfo {author} {\bibfnamefont
  {B.}~\bibnamefont {Kozinsky}},\ }\bibfield  {title} {\enquote {\bibinfo
  {title} {On-the-fly active learning of interpretable {Bayesian} force fields
  for atomistic rare events},}\ }\href {\doibase 10.1038/s41524-020-0283-z}
  {\bibfield  {journal} {\bibinfo  {journal} {npj Comput. Mater.}\ }\textbf
  {\bibinfo {volume} {6}},\ \bibinfo {pages} {20} (\bibinfo {year}
  {2020})}\BibitemShut {NoStop}%
\bibitem [{\citenamefont {Jinnouchi}\ \emph {et~al.}(2020)\citenamefont
  {Jinnouchi}, \citenamefont {Miwa}, \citenamefont {Karsai}, \citenamefont
  {Kresse},\ and\ \citenamefont {Asahi}}]{jinnouchi2020}%
  \BibitemOpen
  \bibfield  {author} {\bibinfo {author} {\bibfnamefont {R.}~\bibnamefont
  {Jinnouchi}}, \bibinfo {author} {\bibfnamefont {K.}~\bibnamefont {Miwa}},
  \bibinfo {author} {\bibfnamefont {F.}~\bibnamefont {Karsai}}, \bibinfo
  {author} {\bibfnamefont {G.}~\bibnamefont {Kresse}}, \ and\ \bibinfo {author}
  {\bibfnamefont {R.}~\bibnamefont {Asahi}},\ }\bibfield  {title} {\enquote
  {\bibinfo {title} {On-the-{Fly} {Active} {Learning} of {Interatomic}
  {Potentials} for {Large}-{Scale} {Atomistic} {Simulations}},}\ }\href
  {\doibase 10.1021/acs.jpclett.0c01061} {\bibfield  {journal} {\bibinfo
  {journal} {J. Phys. Chem. Lett.}\ }\textbf {\bibinfo {volume} {11}},\
  \bibinfo {pages} {6946--6955} (\bibinfo {year} {2020})}\BibitemShut {NoStop}%
\bibitem [{\citenamefont {Behler}\ and\ \citenamefont
  {Parrinello}(2007)}]{behler2007}%
  \BibitemOpen
  \bibfield  {author} {\bibinfo {author} {\bibfnamefont {J.}~\bibnamefont
  {Behler}}\ and\ \bibinfo {author} {\bibfnamefont {M.}~\bibnamefont
  {Parrinello}},\ }\bibfield  {title} {\enquote {\bibinfo {title} {Generalized
  {Neural}-{Network} {Representation} of {High}-{Dimensional}
  {Potential}-{Energy} {Surfaces}},}\ }\href {\doibase
  10.1103/PhysRevLett.98.146401} {\bibfield  {journal} {\bibinfo  {journal}
  {Phys. Rev. Lett.}\ }\textbf {\bibinfo {volume} {98}},\ \bibinfo {pages}
  {146401} (\bibinfo {year} {2007})}\BibitemShut {NoStop}%
\bibitem [{\citenamefont {Behler}(2011)}]{behler2011}%
  \BibitemOpen
  \bibfield  {author} {\bibinfo {author} {\bibfnamefont {J.}~\bibnamefont
  {Behler}},\ }\bibfield  {title} {\enquote {\bibinfo {title} {Atom-centered
  symmetry functions for constructing high-dimensional neural network
  potentials},}\ }\href {\doibase 10.1063/1.3553717} {\bibfield  {journal}
  {\bibinfo  {journal} {J. Chem. Phys.}\ }\textbf {\bibinfo {volume} {134}},\
  \bibinfo {pages} {074106} (\bibinfo {year} {2011})}\BibitemShut {NoStop}%
\bibitem [{\citenamefont {Behler}(2015)}]{behler2015}%
  \BibitemOpen
  \bibfield  {author} {\bibinfo {author} {\bibfnamefont {J.}~\bibnamefont
  {Behler}},\ }\bibfield  {title} {\enquote {\bibinfo {title} {Constructing
  high-dimensional neural network potentials: {A} tutorial review},}\ }\href
  {\doibase 10.1002/qua.24890} {\bibfield  {journal} {\bibinfo  {journal} {Int.
  J. Quantum Chem.}\ }\textbf {\bibinfo {volume} {115}},\ \bibinfo {pages}
  {1032--1050} (\bibinfo {year} {2015})}\BibitemShut {NoStop}%
\bibitem [{\citenamefont {Watanabe}\ \emph {et~al.}(2020)\citenamefont
  {Watanabe}, \citenamefont {Li}, \citenamefont {Jeong}, \citenamefont {Lee},
  \citenamefont {Shimizu}, \citenamefont {Mimanitani}, \citenamefont {Ando},\
  and\ \citenamefont {Han}}]{Watanabe2020}%
  \BibitemOpen
  \bibfield  {author} {\bibinfo {author} {\bibfnamefont {S.}~\bibnamefont
  {Watanabe}}, \bibinfo {author} {\bibfnamefont {W.}~\bibnamefont {Li}},
  \bibinfo {author} {\bibfnamefont {W.}~\bibnamefont {Jeong}}, \bibinfo
  {author} {\bibfnamefont {D.}~\bibnamefont {Lee}}, \bibinfo {author}
  {\bibfnamefont {K.}~\bibnamefont {Shimizu}}, \bibinfo {author} {\bibfnamefont
  {E.}~\bibnamefont {Mimanitani}}, \bibinfo {author} {\bibfnamefont
  {Y.}~\bibnamefont {Ando}}, \ and\ \bibinfo {author} {\bibfnamefont
  {S.}~\bibnamefont {Han}},\ }\bibfield  {title} {\enquote {\bibinfo {title}
  {High-dimensional neural network atomic potentials for examining energy
  materials: some recent simulations},}\ }\href {\doibase
  10.1088/2515-7655/abc7f3} {\bibfield  {journal} {\bibinfo  {journal} {J.
  Phys.: Energy}\ }\textbf {\bibinfo {volume} {3}},\ \bibinfo {pages} {012003}
  (\bibinfo {year} {2020})}\BibitemShut {NoStop}%
\bibitem [{\citenamefont {Bart\'ok}, \citenamefont {Kondor},\ and\
  \citenamefont {Cs\'anyi}(2013)}]{Bartok2013}%
  \BibitemOpen
  \bibfield  {author} {\bibinfo {author} {\bibfnamefont {A.~P.}\ \bibnamefont
  {Bart\'ok}}, \bibinfo {author} {\bibfnamefont {R.}~\bibnamefont {Kondor}}, \
  and\ \bibinfo {author} {\bibfnamefont {G.}~\bibnamefont {Cs\'anyi}},\
  }\bibfield  {title} {\enquote {\bibinfo {title} {On representing chemical
  environments},}\ }\href {\doibase 10.1103/PhysRevB.87.184115} {\bibfield
  {journal} {\bibinfo  {journal} {Phys. Rev. B}\ }\textbf {\bibinfo {volume}
  {87}},\ \bibinfo {pages} {184115} (\bibinfo {year} {2013})}\BibitemShut
  {NoStop}%
\bibitem [{\citenamefont {Zhang}\ \emph
  {et~al.}(2018{\natexlab{a}})\citenamefont {Zhang}, \citenamefont {Han},
  \citenamefont {Wang}, \citenamefont {Car},\ and\ \citenamefont
  {E}}]{Zhang2018b}%
  \BibitemOpen
  \bibfield  {author} {\bibinfo {author} {\bibfnamefont {L.}~\bibnamefont
  {Zhang}}, \bibinfo {author} {\bibfnamefont {J.}~\bibnamefont {Han}}, \bibinfo
  {author} {\bibfnamefont {H.}~\bibnamefont {Wang}}, \bibinfo {author}
  {\bibfnamefont {R.}~\bibnamefont {Car}}, \ and\ \bibinfo {author}
  {\bibfnamefont {W.}~\bibnamefont {E}},\ }\bibfield  {title} {\enquote
  {\bibinfo {title} {Deep {Potential} {Molecular} {Dynamics}: {A} {Scalable}
  {Model} with the {Accuracy} of {Quantum} {Mechanics}},}\ }\href {\doibase
  10.1103/PhysRevLett.120.143001} {\bibfield  {journal} {\bibinfo  {journal}
  {Phys. Rev. Lett.}\ }\textbf {\bibinfo {volume} {120}},\ \bibinfo {pages}
  {143001} (\bibinfo {year} {2018}{\natexlab{a}})}\BibitemShut {NoStop}%
\bibitem [{\citenamefont {Zhang}\ \emph
  {et~al.}(2018{\natexlab{b}})\citenamefont {Zhang}, \citenamefont {Han},
  \citenamefont {Wang}, \citenamefont {Saidi}, \citenamefont {Car},\ and\
  \citenamefont {E}}]{Zhang2018c}%
  \BibitemOpen
  \bibfield  {author} {\bibinfo {author} {\bibfnamefont {L.}~\bibnamefont
  {Zhang}}, \bibinfo {author} {\bibfnamefont {J.}~\bibnamefont {Han}}, \bibinfo
  {author} {\bibfnamefont {H.}~\bibnamefont {Wang}}, \bibinfo {author}
  {\bibfnamefont {W.}~\bibnamefont {Saidi}}, \bibinfo {author} {\bibfnamefont
  {R.}~\bibnamefont {Car}}, \ and\ \bibinfo {author} {\bibfnamefont
  {W.}~\bibnamefont {E}},\ }\bibfield  {title} {\enquote {\bibinfo {title}
  {End-to-end {Symmetry} {Preserving} {Inter}-atomic {Potential} {Energy}
  {Model} for {Finite} and {Extended} {Systems}},}\ }\href
  {https://proceedings.neurips.cc/paper/2018/hash/e2ad76f2326fbc6b56a45a56c59fafdb-Abstract.html}
  {\bibfield  {journal} {\bibinfo  {journal} {Adv. Neural. Inf. Process Syst.}\
  }\textbf {\bibinfo {volume} {31}},\ \bibinfo {pages} {4436--4446} (\bibinfo
  {year} {2018}{\natexlab{b}})}\BibitemShut {NoStop}%
\bibitem [{\citenamefont {Artrith}, \citenamefont {Urban},\ and\ \citenamefont
  {Ceder}(2017)}]{artrith2017}%
  \BibitemOpen
  \bibfield  {author} {\bibinfo {author} {\bibfnamefont {N.}~\bibnamefont
  {Artrith}}, \bibinfo {author} {\bibfnamefont {A.}~\bibnamefont {Urban}}, \
  and\ \bibinfo {author} {\bibfnamefont {G.}~\bibnamefont {Ceder}},\ }\bibfield
   {title} {\enquote {\bibinfo {title} {Efficient and accurate machine-learning
  interpolation of atomic energies in compositions with many species},}\ }\href
  {\doibase 10.1103/PhysRevB.96.014112} {\bibfield  {journal} {\bibinfo
  {journal} {Phys. Rev. B}\ }\textbf {\bibinfo {volume} {96}},\ \bibinfo
  {pages} {014112} (\bibinfo {year} {2017})}\BibitemShut {NoStop}%
\bibitem [{\citenamefont {Li}, \citenamefont {Ando},\ and\ \citenamefont
  {Watanabe}(2017)}]{Li2017_Cu}%
  \BibitemOpen
  \bibfield  {author} {\bibinfo {author} {\bibfnamefont {W.}~\bibnamefont
  {Li}}, \bibinfo {author} {\bibfnamefont {Y.}~\bibnamefont {Ando}}, \ and\
  \bibinfo {author} {\bibfnamefont {S.}~\bibnamefont {Watanabe}},\ }\bibfield
  {title} {\enquote {\bibinfo {title} {Cu {Diffusion} in {Amorphous}
  {Ta}$_2${O}$_5$ {Studied} with a {Simplified} {Neural} {Network}
  {Potential}},}\ }\href {\doibase 10.7566/JPSJ.86.104004} {\bibfield
  {journal} {\bibinfo  {journal} {J. Phys. Soc. Jpn.}\ }\textbf {\bibinfo
  {volume} {86}},\ \bibinfo {pages} {104004} (\bibinfo {year}
  {2017})}\BibitemShut {NoStop}%
\bibitem [{\citenamefont {Yuge}\ \emph {et~al.}(2007)\citenamefont {Yuge},
  \citenamefont {Seko}, \citenamefont {Kuwabara}, \citenamefont {Oba},\ and\
  \citenamefont {Tanaka}}]{Yuge2007}%
  \BibitemOpen
  \bibfield  {author} {\bibinfo {author} {\bibfnamefont {K.}~\bibnamefont
  {Yuge}}, \bibinfo {author} {\bibfnamefont {A.}~\bibnamefont {Seko}}, \bibinfo
  {author} {\bibfnamefont {A.}~\bibnamefont {Kuwabara}}, \bibinfo {author}
  {\bibfnamefont {F.}~\bibnamefont {Oba}}, \ and\ \bibinfo {author}
  {\bibfnamefont {I.}~\bibnamefont {Tanaka}},\ }\bibfield  {title} {\enquote
  {\bibinfo {title} {Ordering and segregation of a {Cu}$_{75}${Pt}$_{25}$ (111)
  surface: {A} first-principles cluster expansion study},}\ }\href {\doibase
  10.1103/PhysRevB.76.045407} {\bibfield  {journal} {\bibinfo  {journal} {Phys.
  Rev. B}\ }\textbf {\bibinfo {volume} {76}},\ \bibinfo {pages} {045407}
  (\bibinfo {year} {2007})}\BibitemShut {NoStop}%
\bibitem [{\citenamefont {Ji}\ and\ \citenamefont {Jung}(2017)}]{Ji2017}%
  \BibitemOpen
  \bibfield  {author} {\bibinfo {author} {\bibfnamefont {H.}~\bibnamefont
  {Ji}}\ and\ \bibinfo {author} {\bibfnamefont {Y.}~\bibnamefont {Jung}},\
  }\bibfield  {title} {\enquote {\bibinfo {title} {Artificial neural network
  for the configuration problem in solids},}\ }\href {\doibase
  10.1063/1.4974928} {\bibfield  {journal} {\bibinfo  {journal} {J. Chem.
  Phys.}\ }\textbf {\bibinfo {volume} {146}},\ \bibinfo {pages} {064103}
  (\bibinfo {year} {2017})}\BibitemShut {NoStop}%
\bibitem [{\citenamefont {Natarajan}\ and\ \citenamefont {Van~der
  Ven}(2018)}]{Natarajan2018}%
  \BibitemOpen
  \bibfield  {author} {\bibinfo {author} {\bibfnamefont {A.~R.}\ \bibnamefont
  {Natarajan}}\ and\ \bibinfo {author} {\bibfnamefont {A.}~\bibnamefont
  {Van~der Ven}},\ }\bibfield  {title} {\enquote {\bibinfo {title}
  {Machine-learning the configurational energy of multicomponent crystalline
  solids},}\ }\href {\doibase 10.1038/s41524-018-0110-y} {\bibfield  {journal}
  {\bibinfo  {journal} {npj Comput. Mater.}\ }\textbf {\bibinfo {volume} {4}},\
  \bibinfo {pages} {56} (\bibinfo {year} {2018})}\BibitemShut {NoStop}%
\bibitem [{\citenamefont {Shapeev}(2017)}]{Shapeev2017}%
  \BibitemOpen
  \bibfield  {author} {\bibinfo {author} {\bibfnamefont {A.}~\bibnamefont
  {Shapeev}},\ }\bibfield  {title} {\enquote {\bibinfo {title} {Accurate
  representation of formation energies of crystalline alloys with many
  components},}\ }\href {\doibase 10.1016/j.commatsci.2017.07.010} {\bibfield
  {journal} {\bibinfo  {journal} {Comput. Mater. Sci.}\ }\textbf {\bibinfo
  {volume} {139}},\ \bibinfo {pages} {26--30} (\bibinfo {year}
  {2017})}\BibitemShut {NoStop}%
\bibitem [{\citenamefont {Biagioni}\ and\ \citenamefont
  {Pasero}(2014)}]{Biagioni2014}%
  \BibitemOpen
  \bibfield  {author} {\bibinfo {author} {\bibfnamefont {C.}~\bibnamefont
  {Biagioni}}\ and\ \bibinfo {author} {\bibfnamefont {M.}~\bibnamefont
  {Pasero}},\ }\bibfield  {title} {\enquote {\bibinfo {title} {The systematics
  of the spinel-type minerals: {An} overview},}\ }\href {\doibase
  10.2138/am.2014.4816} {\bibfield  {journal} {\bibinfo  {journal} {Am.
  Mineral.}\ }\textbf {\bibinfo {volume} {99}},\ \bibinfo {pages} {1254--1264}
  (\bibinfo {year} {2014})}\BibitemShut {NoStop}%
\bibitem [{\citenamefont {Thackeray}(2021)}]{Thackeray2021}%
  \BibitemOpen
  \bibfield  {author} {\bibinfo {author} {\bibfnamefont {M.~M.}\ \bibnamefont
  {Thackeray}},\ }\bibfield  {title} {\enquote {\bibinfo {title} {Exploiting
  the {Spinel} {Structure} for {Li}-ion {Battery} {Applications}: {A} {Tribute}
  to {John} {B}. {Goodenough}},}\ }\href {\doibase 10.1002/aenm.202001117}
  {\bibfield  {journal} {\bibinfo  {journal} {Adv. Energy Mater.}\ }\textbf
  {\bibinfo {volume} {11}},\ \bibinfo {pages} {2001117} (\bibinfo {year}
  {2021})}\BibitemShut {NoStop}%
\bibitem [{\citenamefont {Narang}\ and\ \citenamefont
  {Pubby}(2021)}]{Narang2021}%
  \BibitemOpen
  \bibfield  {author} {\bibinfo {author} {\bibfnamefont {S.~B.}\ \bibnamefont
  {Narang}}\ and\ \bibinfo {author} {\bibfnamefont {K.}~\bibnamefont {Pubby}},\
  }\bibfield  {title} {\enquote {\bibinfo {title} {Nickel {Spinel} {Ferrites}:
  {A} review},}\ }\href {\doibase 10.1016/j.jmmm.2020.167163} {\bibfield
  {journal} {\bibinfo  {journal} {J. Magn. Magn. Mater.}\ }\textbf {\bibinfo
  {volume} {519}},\ \bibinfo {pages} {167163} (\bibinfo {year}
  {2021})}\BibitemShut {NoStop}%
\bibitem [{\citenamefont {Seko}\ \emph {et~al.}(2006)\citenamefont {Seko},
  \citenamefont {Yuge}, \citenamefont {Oba}, \citenamefont {Kuwabara},\ and\
  \citenamefont {Tanaka}}]{Seko2006}%
  \BibitemOpen
  \bibfield  {author} {\bibinfo {author} {\bibfnamefont {A.}~\bibnamefont
  {Seko}}, \bibinfo {author} {\bibfnamefont {K.}~\bibnamefont {Yuge}}, \bibinfo
  {author} {\bibfnamefont {F.}~\bibnamefont {Oba}}, \bibinfo {author}
  {\bibfnamefont {A.}~\bibnamefont {Kuwabara}}, \ and\ \bibinfo {author}
  {\bibfnamefont {I.}~\bibnamefont {Tanaka}},\ }\bibfield  {title} {\enquote
  {\bibinfo {title} {Prediction of ground-state structures and order-disorder
  phase transitions in {II}-{III} spinel oxides: {A} combined cluster-expansion
  method and first-principles study},}\ }\href {\doibase
  10.1103/PhysRevB.73.184117} {\bibfield  {journal} {\bibinfo  {journal} {Phys.
  Rev. B}\ }\textbf {\bibinfo {volume} {73}},\ \bibinfo {pages} {184117}
  (\bibinfo {year} {2006})}\BibitemShut {NoStop}%
\bibitem [{\citenamefont {Kresse}\ and\ \citenamefont
  {Furthm\"{u}ller}(1996{\natexlab{a}})}]{Kresse1996}%
  \BibitemOpen
  \bibfield  {author} {\bibinfo {author} {\bibfnamefont {G.}~\bibnamefont
  {Kresse}}\ and\ \bibinfo {author} {\bibfnamefont {J.}~\bibnamefont
  {Furthm\"{u}ller}},\ }\bibfield  {title} {\enquote {\bibinfo {title}
  {{Efficient iterative schemes for ab initio total-energy calculations using a
  plane-wave basis set}},}\ }\href@noop {} {\bibfield  {journal} {\bibinfo
  {journal} {Phys. Rev. B}\ }\textbf {\bibinfo {volume} {54}},\ \bibinfo
  {pages} {11169--11186} (\bibinfo {year} {1996}{\natexlab{a}})}\BibitemShut
  {NoStop}%
\bibitem [{\citenamefont {Kresse}\ and\ \citenamefont
  {Furthm\"{u}ller}(1996{\natexlab{b}})}]{Kresse1996a}%
  \BibitemOpen
  \bibfield  {author} {\bibinfo {author} {\bibfnamefont {G.}~\bibnamefont
  {Kresse}}\ and\ \bibinfo {author} {\bibfnamefont {J.}~\bibnamefont
  {Furthm\"{u}ller}},\ }\href@noop {} {\bibfield  {journal} {\bibinfo
  {journal} {Comp. Mater. Sci.}\ }\textbf {\bibinfo {volume} {6}},\ \bibinfo
  {pages} {15} (\bibinfo {year} {1996}{\natexlab{b}})}\BibitemShut {NoStop}%
\bibitem [{\citenamefont {Bl\"{o}chl}(1994)}]{Blochl1994}%
  \BibitemOpen
  \bibfield  {author} {\bibinfo {author} {\bibfnamefont {P.~E.}\ \bibnamefont
  {Bl\"{o}chl}},\ }\bibfield  {title} {\enquote {\bibinfo {title} {{Projector
  augmented-wave method}},}\ }\href@noop {} {\bibfield  {journal} {\bibinfo
  {journal} {Phys. Rev. B}\ }\textbf {\bibinfo {volume} {50}},\ \bibinfo
  {pages} {17953--17979} (\bibinfo {year} {1994})}\BibitemShut {NoStop}%
\bibitem [{\citenamefont {Perdew}, \citenamefont {Burke},\ and\ \citenamefont
  {Ernzerhof}(1996)}]{Perdew1996}%
  \BibitemOpen
  \bibfield  {author} {\bibinfo {author} {\bibfnamefont {J.~P.}\ \bibnamefont
  {Perdew}}, \bibinfo {author} {\bibfnamefont {K.}~\bibnamefont {Burke}}, \
  and\ \bibinfo {author} {\bibfnamefont {M.}~\bibnamefont {Ernzerhof}},\
  }\bibfield  {title} {\enquote {\bibinfo {title} {{Generalized Gradient
  Approximation Made Simple}},}\ }\href@noop {} {\bibfield  {journal} {\bibinfo
   {journal} {Phys. Rev. Lett.}\ }\textbf {\bibinfo {volume} {77}},\ \bibinfo
  {pages} {3865--3868} (\bibinfo {year} {1996})}\BibitemShut {NoStop}%
\bibitem [{\citenamefont {Artrith}\ and\ \citenamefont
  {Urban}(2016)}]{artrith2016}%
  \BibitemOpen
  \bibfield  {author} {\bibinfo {author} {\bibfnamefont {N.}~\bibnamefont
  {Artrith}}\ and\ \bibinfo {author} {\bibfnamefont {A.}~\bibnamefont
  {Urban}},\ }\bibfield  {title} {\enquote {\bibinfo {title} {An implementation
  of artificial neural-network potentials for atomistic materials simulations:
  {Performance} for {TiO2}},}\ }\href {\doibase
  10.1016/j.commatsci.2015.11.047} {\bibfield  {journal} {\bibinfo  {journal}
  {Comput. Mater. Sci.}\ }\textbf {\bibinfo {volume} {114}},\ \bibinfo {pages}
  {135--150} (\bibinfo {year} {2016})}\BibitemShut {NoStop}%
\bibitem [{Note1()}]{Note1}%
  \BibitemOpen
  \bibinfo {note} {We have 16 coefficients for RDF and 4 for ADF resulting in
  20 descriptors, and this is multiplied by 2 since we have separate RDFs/ADFs
  for composition and structure.}\BibitemShut {Stop}%
\bibitem [{\citenamefont {Liu}\ \emph {et~al.}(2021)\citenamefont {Liu},
  \citenamefont {Zhang}, \citenamefont {Yin}, \citenamefont {Bi}, \citenamefont
  {Eisenbach},\ and\ \citenamefont {Wang}}]{Liu2021}%
  \BibitemOpen
  \bibfield  {author} {\bibinfo {author} {\bibfnamefont {X.}~\bibnamefont
  {Liu}}, \bibinfo {author} {\bibfnamefont {J.}~\bibnamefont {Zhang}}, \bibinfo
  {author} {\bibfnamefont {J.}~\bibnamefont {Yin}}, \bibinfo {author}
  {\bibfnamefont {S.}~\bibnamefont {Bi}}, \bibinfo {author} {\bibfnamefont
  {M.}~\bibnamefont {Eisenbach}}, \ and\ \bibinfo {author} {\bibfnamefont
  {Y.}~\bibnamefont {Wang}},\ }\bibfield  {title} {\enquote {\bibinfo {title}
  {Monte {Carlo} simulation of order-disorder transition in refractory high
  entropy alloys: {A} data-driven approach},}\ }\href {\doibase
  10.1016/j.commatsci.2020.110135} {\bibfield  {journal} {\bibinfo  {journal}
  {Comput. Mater. Sci.}\ }\textbf {\bibinfo {volume} {187}},\ \bibinfo {pages}
  {110135} (\bibinfo {year} {2021})}\BibitemShut {NoStop}%
\bibitem [{Note2()}]{Note2}%
  \BibitemOpen
  \bibinfo {note} {The splitting of samples between higher and lower energies
  with few samples in the interval of e.g., $\sim $ 50--75 meV/cation for
  MgAl$_2$O$_4$ originates from our sampling procedure: we always sampled the
  first step in the RXMC calculations which is generated randomly, and most of
  the replicas relax quickly to lower energy configurations before the next
  sampling step.}\BibitemShut {Stop}%
\bibitem [{\citenamefont {Machatschki}(1932)}]{Machatschki1932}%
  \BibitemOpen
  \bibfield  {author} {\bibinfo {author} {\bibfnamefont {F.}~\bibnamefont
  {Machatschki}},\ }\bibfield  {title} {\enquote {\bibinfo {title} {Der
  {Magnesium}-{Gallium}-{Spinell}},}\ }\href {\doibase
  doi:10.1524/zkri.1932.82.1.348} {\bibfield  {journal} {\bibinfo  {journal}
  {Zeitschrift für Kristallographie - Crystalline Materials}\ }\textbf
  {\bibinfo {volume} {82}},\ \bibinfo {pages} {348--354} (\bibinfo {year}
  {1932})}\BibitemShut {NoStop}%
\bibitem [{\citenamefont {M.}(1957)}]{Huber1957}%
  \BibitemOpen
  \bibfield  {author} {\bibinfo {author} {\bibfnamefont {H.}~\bibnamefont
  {M.}},\ }\href@noop {} {\bibfield  {journal} {\bibinfo  {journal} {C. R.
  Acad. Sci., Paris}\ }\textbf {\bibinfo {volume} {244}},\ \bibinfo {pages}
  {2524} (\bibinfo {year} {1957})}\BibitemShut {NoStop}%
\bibitem [{\citenamefont {Schmalzried}(1961)}]{Schmalzried1961}%
  \BibitemOpen
  \bibfield  {author} {\bibinfo {author} {\bibfnamefont {H.}~\bibnamefont
  {Schmalzried}},\ }\bibfield  {title} {\enquote {\bibinfo {title}
  {R\"ontgenographische {Untersuchung} der {Kationenverteilung} in
  {Spinellphasen}},}\ }\href {\doibase doi:10.1524/zpch.1961.28.3_4.203}
  {\bibfield  {journal} {\bibinfo  {journal} {Z. Phys. Chem.}\ }\textbf
  {\bibinfo {volume} {28}},\ \bibinfo {pages} {203--219} (\bibinfo {year}
  {1961})}\BibitemShut {NoStop}%
\bibitem [{\citenamefont {Weidenborner}, \citenamefont {Stemple},\ and\
  \citenamefont {Okaya}(1966)}]{Weidenborner1966}%
  \BibitemOpen
  \bibfield  {author} {\bibinfo {author} {\bibfnamefont {J.~E.}\ \bibnamefont
  {Weidenborner}}, \bibinfo {author} {\bibfnamefont {N.~R.}\ \bibnamefont
  {Stemple}}, \ and\ \bibinfo {author} {\bibfnamefont {Y.}~\bibnamefont
  {Okaya}},\ }\bibfield  {title} {\enquote {\bibinfo {title} {Cation
  distribution and oxygen parameter in magnesium gallate, {MgGa}
  $_{\textrm{2}}$ {O} $_{\textrm{4}}$},}\ }\href {\doibase
  10.1107/S0365110X66001816} {\bibfield  {journal} {\bibinfo  {journal} {Acta
  Crystallogr.}\ }\textbf {\bibinfo {volume} {20}},\ \bibinfo {pages}
  {761--764} (\bibinfo {year} {1966})}\BibitemShut {NoStop}%
\bibitem [{\citenamefont {Barth}\ and\ \citenamefont
  {Posnjak}(1932)}]{Barth1932}%
  \BibitemOpen
  \bibfield  {author} {\bibinfo {author} {\bibfnamefont {T.~F.~W.}\
  \bibnamefont {Barth}}\ and\ \bibinfo {author} {\bibfnamefont
  {E.}~\bibnamefont {Posnjak}},\ }\bibfield  {title} {\enquote {\bibinfo
  {title} {Spinel structures: with and without variate atom equipoints},}\
  }\href {\doibase doi:10.1524/zkri.1932.82.1.325} {\bibfield  {journal}
  {\bibinfo  {journal} {Zeitschrift für Kristallographie - Crystalline
  Materials}\ }\textbf {\bibinfo {volume} {82}},\ \bibinfo {pages} {325--341}
  (\bibinfo {year} {1932})}\BibitemShut {NoStop}%
\bibitem [{\citenamefont {Dai}\ \emph {et~al.}(2020)\citenamefont {Dai},
  \citenamefont {Wen}, \citenamefont {Sun}, \citenamefont {Xiang},\ and\
  \citenamefont {Zhou}}]{Dai2020}%
  \BibitemOpen
  \bibfield  {author} {\bibinfo {author} {\bibfnamefont {F.-Z.}\ \bibnamefont
  {Dai}}, \bibinfo {author} {\bibfnamefont {B.}~\bibnamefont {Wen}}, \bibinfo
  {author} {\bibfnamefont {Y.}~\bibnamefont {Sun}}, \bibinfo {author}
  {\bibfnamefont {H.}~\bibnamefont {Xiang}}, \ and\ \bibinfo {author}
  {\bibfnamefont {Y.}~\bibnamefont {Zhou}},\ }\bibfield  {title} {\enquote
  {\bibinfo {title} {Theoretical prediction on thermal and mechanical
  properties of high entropy
  ({Zr}$_{0.2}${Hf}$_{0.2}${Ti}$_{0.2}${Nb}$_{0.2}${Ta}$_{0.2}$){C} by deep
  learning potential},}\ }\href {\doibase 10.1016/j.jmst.2020.01.005}
  {\bibfield  {journal} {\bibinfo  {journal} {J. Mater. Sci. Technol.}\
  }\textbf {\bibinfo {volume} {43}},\ \bibinfo {pages} {168--174} (\bibinfo
  {year} {2020})}\BibitemShut {NoStop}%
\bibitem [{\citenamefont {Artrith}, \citenamefont {Morawietz},\ and\
  \citenamefont {Behler}(2011)}]{Artrith2011}%
  \BibitemOpen
  \bibfield  {author} {\bibinfo {author} {\bibfnamefont {N.}~\bibnamefont
  {Artrith}}, \bibinfo {author} {\bibfnamefont {T.}~\bibnamefont {Morawietz}},
  \ and\ \bibinfo {author} {\bibfnamefont {J.}~\bibnamefont {Behler}},\
  }\bibfield  {title} {\enquote {\bibinfo {title} {High-dimensional
  neural-network potentials for multicomponent systems: {Applications} to zinc
  oxide},}\ }\href {\doibase 10.1103/PhysRevB.83.153101} {\bibfield  {journal}
  {\bibinfo  {journal} {Phys. Rev. B}\ }\textbf {\bibinfo {volume} {83}},\
  \bibinfo {pages} {153101} (\bibinfo {year} {2011})},\ \bibinfo {note}
  {publisher: American Physical Society}\BibitemShut {NoStop}%
\bibitem [{\citenamefont {Iftimie}\ \emph {et~al.}(2000)\citenamefont
  {Iftimie}, \citenamefont {Salahub}, \citenamefont {Wei},\ and\ \citenamefont
  {Schofield}}]{iftimie2000}%
  \BibitemOpen
  \bibfield  {author} {\bibinfo {author} {\bibfnamefont {R.}~\bibnamefont
  {Iftimie}}, \bibinfo {author} {\bibfnamefont {D.}~\bibnamefont {Salahub}},
  \bibinfo {author} {\bibfnamefont {D.}~\bibnamefont {Wei}}, \ and\ \bibinfo
  {author} {\bibfnamefont {J.}~\bibnamefont {Schofield}},\ }\bibfield  {title}
  {\enquote {\bibinfo {title} {Using a classical potential as an efficient
  importance function for sampling from an ab initio potential},}\ }\href
  {\doibase 10.1063/1.1289534} {\bibfield  {journal} {\bibinfo  {journal} {J.
  Chem. Phys.}\ }\textbf {\bibinfo {volume} {113}},\ \bibinfo {pages} {4852}
  (\bibinfo {year} {2000})}\BibitemShut {NoStop}%
\bibitem [{\citenamefont {Het\'enyi}, \citenamefont {Bernacki},\ and\
  \citenamefont {Berne}(2002)}]{Hetenyi2002}%
  \BibitemOpen
  \bibfield  {author} {\bibinfo {author} {\bibfnamefont {B.}~\bibnamefont
  {Het\'enyi}}, \bibinfo {author} {\bibfnamefont {K.}~\bibnamefont {Bernacki}},
  \ and\ \bibinfo {author} {\bibfnamefont {B.~J.}\ \bibnamefont {Berne}},\
  }\bibfield  {title} {\enquote {\bibinfo {title} {Multiple “time step”
  {Monte} {Carlo}},}\ }\href {\doibase 10.1063/1.1512645} {\bibfield  {journal}
  {\bibinfo  {journal} {J. Chem. Phys.}\ }\textbf {\bibinfo {volume} {117}},\
  \bibinfo {pages} {8203--8207} (\bibinfo {year} {2002})},\ \bibinfo {note}
  {publisher: American Institute of Physics}\BibitemShut {NoStop}%
\bibitem [{\citenamefont {Gelb}(2003)}]{gelb2003}%
  \BibitemOpen
  \bibfield  {author} {\bibinfo {author} {\bibfnamefont {L.~D.}\ \bibnamefont
  {Gelb}},\ }\bibfield  {title} {\enquote {\bibinfo {title} {Monte {Carlo}
  simulations using sampling from an approximate potential},}\ }\href {\doibase
  10.1063/1.1563597} {\bibfield  {journal} {\bibinfo  {journal} {J. Chem.
  Phys.}\ }\textbf {\bibinfo {volume} {118}},\ \bibinfo {pages} {7747--7750}
  (\bibinfo {year} {2003})}\BibitemShut {NoStop}%
\bibitem [{\citenamefont {Leiding}\ and\ \citenamefont
  {Coe}(2016)}]{Leiding2016}%
  \BibitemOpen
  \bibfield  {author} {\bibinfo {author} {\bibfnamefont {J.}~\bibnamefont
  {Leiding}}\ and\ \bibinfo {author} {\bibfnamefont {J.~D.}\ \bibnamefont
  {Coe}},\ }\bibfield  {title} {\enquote {\bibinfo {title} {{Reactive Monte
  Carlo sampling with an {\it ab initio} potential}},}\ }\href {\doibase
  10.1063/1.4948303} {\bibfield  {journal} {\bibinfo  {journal} {J. Chem.
  Phys.}\ }\textbf {\bibinfo {volume} {144}},\ \bibinfo {pages} {174109}
  (\bibinfo {year} {2016})}\BibitemShut {NoStop}%
\bibitem [{\citenamefont {Nagai}, \citenamefont {Okumura},\ and\ \citenamefont
  {Tanaka}(2020)}]{nagai2020}%
  \BibitemOpen
  \bibfield  {author} {\bibinfo {author} {\bibfnamefont {Y.}~\bibnamefont
  {Nagai}}, \bibinfo {author} {\bibfnamefont {M.}~\bibnamefont {Okumura}}, \
  and\ \bibinfo {author} {\bibfnamefont {A.}~\bibnamefont {Tanaka}},\
  }\bibfield  {title} {\enquote {\bibinfo {title} {Self-learning {Monte}
  {Carlo} method with {Behler}-{Parrinello} neural networks},}\ }\href
  {\doibase 10.1103/PhysRevB.101.115111} {\bibfield  {journal} {\bibinfo
  {journal} {Phys. Rev. B}\ }\textbf {\bibinfo {volume} {101}},\ \bibinfo
  {pages} {115111} (\bibinfo {year} {2020})}\BibitemShut {NoStop}%
\bibitem [{\citenamefont {Nagai}\ \emph {et~al.}(2020)\citenamefont {Nagai},
  \citenamefont {Okumura}, \citenamefont {Kobayashi},\ and\ \citenamefont
  {Shiga}}]{Nagai2020b}%
  \BibitemOpen
  \bibfield  {author} {\bibinfo {author} {\bibfnamefont {Y.}~\bibnamefont
  {Nagai}}, \bibinfo {author} {\bibfnamefont {M.}~\bibnamefont {Okumura}},
  \bibinfo {author} {\bibfnamefont {K.}~\bibnamefont {Kobayashi}}, \ and\
  \bibinfo {author} {\bibfnamefont {M.}~\bibnamefont {Shiga}},\ }\bibfield
  {title} {\enquote {\bibinfo {title} {Self-learning hybrid {Monte} {Carlo}:
  {A} first-principles approach},}\ }\href {\doibase
  10.1103/PhysRevB.102.041124} {\bibfield  {journal} {\bibinfo  {journal}
  {Phys. Rev. B}\ }\textbf {\bibinfo {volume} {102}},\ \bibinfo {pages}
  {041124(R)} (\bibinfo {year} {2020})}\BibitemShut {NoStop}%
\bibitem [{\citenamefont {Zhang}\ \emph
  {et~al.}(2018{\natexlab{c}})\citenamefont {Zhang}, \citenamefont {Grabowski},
  \citenamefont {Hickel},\ and\ \citenamefont {Neugebauer}}]{zhang2018}%
  \BibitemOpen
  \bibfield  {author} {\bibinfo {author} {\bibfnamefont {X.}~\bibnamefont
  {Zhang}}, \bibinfo {author} {\bibfnamefont {B.}~\bibnamefont {Grabowski}},
  \bibinfo {author} {\bibfnamefont {T.}~\bibnamefont {Hickel}}, \ and\ \bibinfo
  {author} {\bibfnamefont {J.}~\bibnamefont {Neugebauer}},\ }\bibfield  {title}
  {\enquote {\bibinfo {title} {Calculating free energies of point defects from
  ab initio},}\ }\href {\doibase 10.1016/j.commatsci.2018.02.042} {\bibfield
  {journal} {\bibinfo  {journal} {Comput. Mater. Sci.}\ }\textbf {\bibinfo
  {volume} {148}},\ \bibinfo {pages} {249--259} (\bibinfo {year}
  {2018}{\natexlab{c}})}\BibitemShut {NoStop}%
\bibitem [{\citenamefont {Wales}\ and\ \citenamefont {Doye}(1997)}]{Wales1997}%
  \BibitemOpen
  \bibfield  {author} {\bibinfo {author} {\bibfnamefont {D.~J.}\ \bibnamefont
  {Wales}}\ and\ \bibinfo {author} {\bibfnamefont {J.~P.}\ \bibnamefont
  {Doye}},\ }\bibfield  {title} {\enquote {\bibinfo {title} {{Global
  optimization by basin-hopping and the lowest energy structures of
  Lennard-Jones clusters containing up to 110 atoms}},}\ }\href@noop {}
  {\bibfield  {journal} {\bibinfo  {journal} {J. Phys. Chem. A}\ }\textbf
  {\bibinfo {volume} {101}},\ \bibinfo {pages} {5111--5116} (\bibinfo {year}
  {1997})}\BibitemShut {NoStop}%
\bibitem [{\citenamefont {Hunter}(2007)}]{Hunter2007}%
  \BibitemOpen
  \bibfield  {author} {\bibinfo {author} {\bibfnamefont {J.~D.}\ \bibnamefont
  {Hunter}},\ }\bibfield  {title} {\enquote {\bibinfo {title} {Matplotlib: A 2d
  graphics environment},}\ }\href {\doibase 10.1109/MCSE.2007.55} {\bibfield
  {journal} {\bibinfo  {journal} {Comput. Sci. Eng.}\ }\textbf {\bibinfo
  {volume} {9}},\ \bibinfo {pages} {90--95} (\bibinfo {year}
  {2007})}\BibitemShut {NoStop}%
\bibitem [{\citenamefont {Waskom}(2021)}]{Waskom2021}%
  \BibitemOpen
  \bibfield  {author} {\bibinfo {author} {\bibfnamefont {M.~L.}\ \bibnamefont
  {Waskom}},\ }\bibfield  {title} {\enquote {\bibinfo {title} {seaborn:
  statistical data visualization},}\ }\href {\doibase 10.21105/joss.03021}
  {\bibfield  {journal} {\bibinfo  {journal} {J. Open Source Softw.}\ }\textbf
  {\bibinfo {volume} {6}},\ \bibinfo {pages} {3021} (\bibinfo {year}
  {2021})}\BibitemShut {NoStop}%
\end{thebibliography}%
\end{document}